\documentclass{elsart1p}
\usepackage{amstext,amsmath,amsfonts,amssymb,amsbsy}
\usepackage{bm}
\usepackage{color,lscape}
\usepackage{graphicx}
\usepackage{multirow}
\usepackage{rotating}

\def\rmBH{\mathrm{BH}}
\def\rmDVCS{\mathrm{DVCS}}
\def\rmI{\mathrm{I}}

\def\CUU{\sigma_{\mathrm{UU}}}
\def\Lumi{\mathcal{L}\,}

\def\ACU{A_\mathrm{C}}
\def\ALU{A_\mathrm{LU}}
\def\ALUDVCS{A_{\mathrm{LU},\rmDVCS}}
\def\ALUI{A_{\mathrm{LU},\rmI}}
\def\CalACU{\mathcal{A}_\mathrm{C}}
\def\CalALUDVCS{\mathcal{A}_{\mathrm{LU}}^{\rmDVCS}}
\def\CalALUI{\mathcal{A}_{\mathrm{LU}}^{\rmI}}

\def\ACUcoh{A_{\mathrm{C},\mathrm{coh}}}
\def\ALUcoh{A_{\mathrm{LU},\rmI,\mathrm{coh}}}

\def\intN{{\mathcal N}}

\def\etal{et al.}

\def\amp2{{\cal T}}

\newcommand{\rd}{\mathrm{d}}
\newcommand{\pms}{\hspace{-0.5mm}\pm\hspace{-0.5mm}}
\newcommand{\de}{{\rm\,d}}

\journal{Nuclear Physics B}

\begin{document}

\begin{frontmatter}

\title{Measurement of azimuthal asymmetries associated with
deeply virtual Compton scattering on an unpolarized deuterium target}

\collab{The HERMES Collaboration}

\author[12,15]{A.~Airapetian}
\author[26]{N.~Akopov}
\author[5]{Z.~Akopov}
\author[6]{M.~Amarian\thanksref{27}}
\author[6]{E.C.~Aschenauer\thanksref{28}}
\author[25]{W.~Augustyniak}
\author[26]{R.~Avakian}
\author[26]{A.~Avetissian}
\author[5]{E.~Avetisyan}
\author[15]{B.~Ball}

% ...
\address[5]{DESY, 22603 Hamburg, Germany}
\address[6]{DESY, 15738 Zeuthen, Germany}
% ...
\address[12]{Physikalisches Institut, Universit\"at Gie{\ss}en, 35392 Gie{\ss}en, Germany}
% ...
\address[15]{Randall Laboratory of Physics, University of Michigan, Ann Arbor, Michigan 48109-1040, USA}
% ...
\address[25]{Andrzej Soltan Institute for Nuclear Studies, 00-689 Warsaw, Poland}
\address[26]{Yerevan Physics Institute, 375036 Yerevan, Armenia}

\begin{abstract}
Azimuthal asymmetries in exclusive electroproduction of a real photon from
an unpolarized deuterium target are measured with respect to beam helicity
and charge. They appear in the distribution of these photons in the
azimuthal angle $\phi$ around the virtual-photon direction, relative to
the lepton scattering plane. The extracted asymmetries are attributed to
either the deeply virtual Compton scattering process or its interference
with the Bethe-Heitler process. They are compared with earlier results on
the proton target. In the measured kinematic region, the beam-charge
asymmetry amplitudes and the leading amplitudes of the beam-helicity
asymmetries on an unpolarized deuteron target are compatible with the
results from unpolarized protons.
\end{abstract}


\end{frontmatter}

\section{Introduction}
\label{sec:Introduction}
Lepton-nucleon scattering experiments have long been an important tool in
the detailed study of  nucleon structure~\cite{BMN:2008}. Two 
complementary approaches have contributed the most to our understanding of 
the nucleon. Elastic lepton-nucleon scattering has been exploited to extract 
nucleon form factors, which reveal how the electromagnetic nucleon 
structure differs from that of a point-like spin-1/2 particle. In another 
approach, Parton Distribution Functions (PDFs) are extracted from Deeply 
Inelastic Scattering (DIS). They represent distributions in the 
longitudinal momentum fraction carried by quarks and gluons in a nucleon 
moving with ``infinite" momentum.
\begin{figure*}[t] 
\centerline{
\includegraphics[width=0.48\columnwidth]{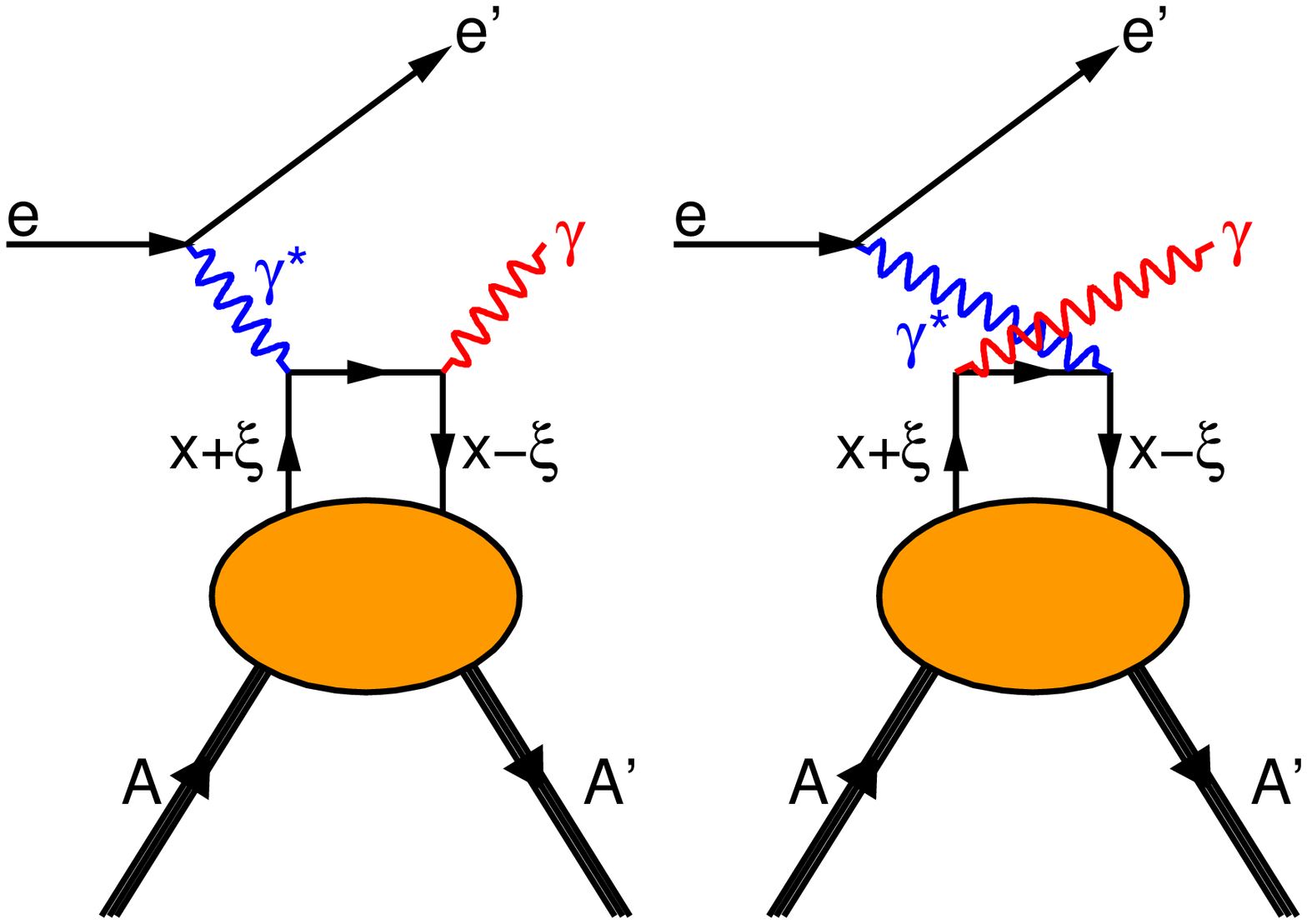}
\includegraphics[width=0.48\columnwidth]{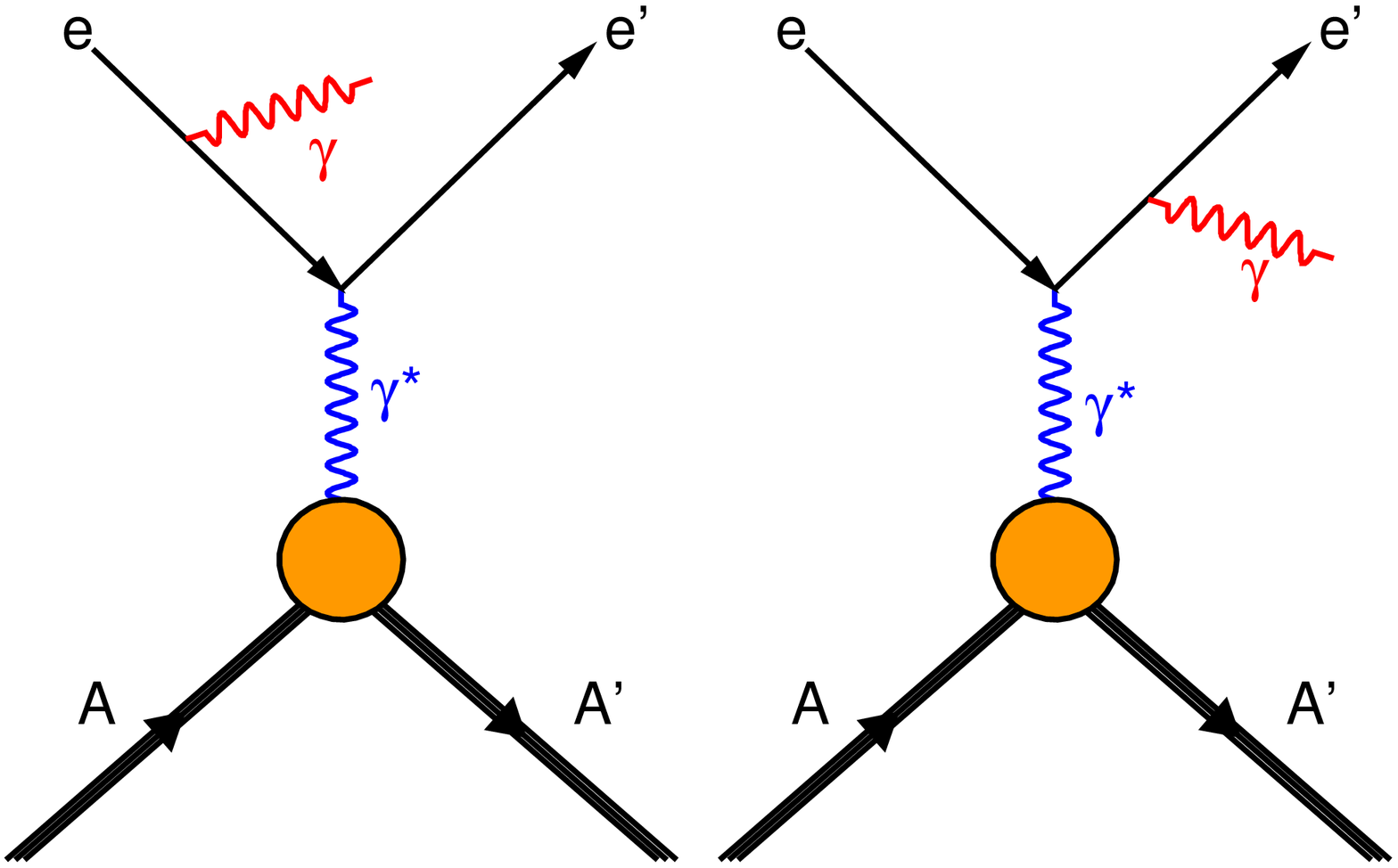}
}
\centerline{{\small (a)}\hspace{.45\textwidth}{\small (b)}}
\caption{Leading order Feynman diagrams for (a) deeply virtual Compton 
scattering and (b) the Bethe-Heitler process.}
\label{fig:DVCSandBH-diagram}
\end{figure*}
PDFs and form factors present only one-dimensional pictures of nucleon 
structure. In recent years, a more comprehensive multi-dimensional 
description of the nucleon has emerged in the framework of Generalized 
Parton Distributions (GPDs)~\cite{Mul94,Rad97,Ji97}. Their dependence on 
three kinematic quantities in addition to their evolution with the hard 
scale of the process carries information on two-parton correlations and 
quark transverse spatial 
distributions~\cite{GPD2,Bur00,GPD3,GPD4,GPD5,GPD6}. GPDs embody PDFs as 
limiting cases, while elastic form factors appear as certain GPD moments.  
Other moments are connected with the total parton angular momentum 
contribution to the nucleon spin via the Ji relation~\cite{Ji97}.

GPDs can be constrained by measurements of hard exclusive leptoproduction 
of a photon or meson in `elastic' processes that leave the target intact. 
In Deeply Virtual Compton Scattering (DVCS), a quark absorbs a hard 
virtual photon, emits an energetic real photon and joins the target 
remnant (see Fig.~\ref{fig:DVCSandBH-diagram} (a)). DVCS is presently the 
only experimentally feasible hard exclusive process for which the effects 
of next-to-leading order~\cite{Bel98,Ji98a,Man98b} and next-to-leading 
twist~\cite{Kiv01a,DVCS2,Fre03b} are under complete theoretical 
control~\cite{BelFreundMul}.

The final state of the DVCS process cannot be experimentally distinguished
from that of the Bethe-Heitler (BH) process, i.e., radiative elastic 
scattering (see Fig.~\ref{fig:DVCSandBH-diagram} (b)). Hence, the two 
processes can interfere. 
Exclusive leptoproduction on a nucleon or nuclear 
target $A$ of a real photon with four-momentum $q'$ is denoted by
\begin{equation}
\label{eq:singlephotonproduction}
e(k) + A(p) \rightarrow e(k^\prime) + A(p^\prime) + \gamma(q^\prime)\,,
\end{equation}
where $k$ $(k^\prime)$ and $p$ $(p^\prime)$ are the four-momenta of the
incoming (outgoing) lepton and target, respectively. Averaged over the 
kinematic acceptance of the HERMES experiment, the BH cross section is 
much larger than that of the DVCS process. However, the BH cross section 
has a much weaker $Q^2$ dependence than the evolution of the DVCS cross 
section~\cite{Ji97}, so that in the HERMES energy range they can become 
comparable near $Q^2=1$\,GeV$^2$, with $-Q^2$ $\equiv$ $q^2$ = 
$(k-k^\prime)^2$.

\begin{figure}
\centerline{
\includegraphics[width=0.5\columnwidth]{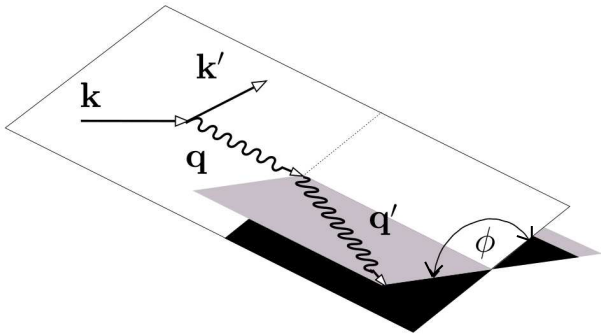}
}
\caption{Definition of the azimuthal angle $\phi$ between the lepton 
scattering and photon production planes. Note that the azimuthal angle 
defined in this work differs from that used in Ref.~\cite{DVCS2}: $\phi = 
\pi - \phi_{{\mbox{\cite{DVCS2}}}}$.}
\label{fig:reaction}
\end{figure}

Even in kinematic conditions where the DVCS process makes only a small 
contribution to the photon production cross section, its interference with 
the BH process provides access to the DVCS amplitudes through measurements 
of cross section asymmetries with respect to the charge and helicity of 
the incident lepton and the polarization of the target. These asymmetries 
appear in the distribution of the real photons in the azimuthal angle 
$\phi$, defined as the angle between the lepton scattering plane, i.e., 
the plane defined by the incoming and outgoing lepton direction and the 
photon production plane spanned by the virtual and real photons (see 
Fig.~\ref{fig:reaction}).
Significant azimuthal beam-helicity asymmetries in hard electroproduction 
of photons on the proton were first reported in 
Refs.~\cite{hermes_bsa:2001,CLAS_bsa:2001}. Later, asymmetries with 
respect to longitudinal~\cite{hermes_tsa:2005,CLAS_tsa:2006} and 
transverse~\cite{hermes_ttsa} target polarization, as well as beam charge 
~\cite{hermes_bca_2006} and, with greater precision, beam 
helicity~\cite{Hall_A:2006,CLAS_bsa:2008,CLAS_bsa:2009,Proton_draft}, were 
also measured on the proton.

Measurements of  azimuthal asymmetries for DVCS on nuclear targets 
\cite{hermes_nuclear} were advocated as a useful source of information 
about partonic behavior in nuclei and  nuclear binding 
forces~\cite{Polyakov:2002yz}. If the target nucleus remains in its ground 
state the process is called coherent, while it is called incoherent if the 
nucleus is broken up. The deuteron is a spin-1 nucleus, with implications 
for DVCS observables for the coherent reactions, which contribute mainly 
at very small values of the momentum transfer to the target. The 
asymmetries from the incoherent process involve mainly hard exclusive 
electroproduction of a photon on the proton. The neutron contribution to 
the yield is typically small due to the suppression of the BH amplitude on 
the neutron by the small elastic electric form factor at low and moderate values 
of the momentum transfer to the target.

This paper reports the  first observation of azimuthal asymmetries with
respect to beam helicity and charge for exclusive electroproduction of a 
real photon from an unpolarized deuterium target ($\vec e^{\;\pm}\,d 
\rightarrow e^{\pm}\,\gamma\,X$). The dependence of these asymmetries on the 
kinematic conditions of the reaction is also presented and certain 
asymmetry amplitudes are compared with the corresponding amplitudes 
obtained on an unpolarized hydrogen target ($\vec e^{\;\pm}\,p\rightarrow 
e^{\pm}\,\gamma\,X $) at HERMES~\cite{Proton_draft}.

\section{GPDs and DVCS} 
\label{sec:GPDsAndDVCS} 
\subsection{Generalized Parton Distributions} 
\label{subsec:GPDs} 
In the generalized Bjorken limit of large $Q^2$ at fixed values of the 
Bjorken scaling variable $x_B=Q^2/(2p\cdot q)$ and small squared four-momentum 
transfer $t = (p-p^\prime)^2$ to the target, the DVCS process can be 
described by the leading (handbag) diagrams in 
Fig.~\ref{fig:DVCSandBH-diagram}(a). Here, the process 
factorizes~\cite{Rad97,Ji98a,Col99} into a hard photon-quark scattering 
part calculable in quantum electrodynamics, and a soft part describing the 
nucleon structure, which can be expressed in terms of 
GPDs~\cite{Mul94,Rad97,Ji97}.  

Like PDFs, GPDs depend on $x$ and on the factorization scale $Q^2$. In 
addition, GPDs depend on a skewness variable $\xi$ and the Mandelstam 
variable $t$. The skewness $\xi$ represents half the difference in the 
longitudinal momentum fractions of the quark before and after the 
scattering, while $x$ is their mean value (following the convention of 
Ref.~\cite{Ji97}). In leading order, $\xi$ is directly accessible as it is 
related to the Bjorken scaling variable $x_B$ by $\xi \simeq x_B/(2-x_B)$. 
In contrast, $x$ is not directly accessible in DVCS, and some observables 
appear as $x$-convolutions of GPDs. Hence $x$ plays a role different from 
that of $x_B$ in inclusive DIS. GPDs evolve logarithmically with $Q^2$ in 
analogy with PDFs~\cite{Mul94,Rad97,Ji97,Blu97}. This dependence on $Q^2$ 
is omitted for simplicity in the following.

DVCS on spin-1/2 targets, such as nucleons, is described by four 
leading-twist quark-chirality conserving GPDs for each quark flavour $q$
(and also for the gluon $g$), namely the GPDs $H^q$, $E^q$, 
$\widetilde{H}^q$ and $\widetilde{E}^q$~\cite{DVCS2}. The GPDs $H^q$ and 
$E^q$ are quark-helicity averaged whereas $\widetilde{H}^q$ and 
$\widetilde{E}^q$ are quark-helicity dependent. The GPDs $H^q$ and 
$\widetilde{H}^q$ conserve nucleon-helicity while $E^q$ and 
$\widetilde{E}^q$ are associated with a helicity flip of the 
nucleon. In contrast, the coherent process on spin-1 nuclei, such as the 
deuteron, requires nine GPDs~\cite{Berger:2001zb} --- $H^q_1$, $H^q_2$, 
$H^q_3$, $H^q_4$, $H^q_5$, $\widetilde{H}^q_1$, $\widetilde{H}^q_2$, 
$\widetilde{H}^q_3$ and $\widetilde{H}^q_4$ --- to describe all DVCS 
observables. In the forward limit of vanishing momentum difference between 
the initial and final hadronic state ($t \rightarrow 0$ and $\xi 
\rightarrow 0$), the GPD $H^q(x,0,0)$ reduces to $f_1^q(x)$, the quark 
number density distribution, and $\widetilde{H}^q(x,0,0)$ reduces to 
$g_1^q(x)$, the quark helicity distribution. Similarly, for spin-1 targets 
the GPDs $H_1$, $\widetilde{H}_1$ and $H_5$ reduce to the following parton 
densities in the forward limit:
\begin{alignat}{2}
H^q_1(x,0,0)  &=   \frac{q^1(x) + q^{-1}(x) + q^0(x)}{3} &\equiv 
f_1^q(x)\,,
\label{eq:forward1} \\
\widetilde{H}^q_1(x,0,0) &=   q^1_\rightarrow (x)  - 
q^{-1}_\rightarrow (x) &\equiv  g_1^q(x)\,, 
\label{eq:forward2} \\
H^q_5(x,0,0)   &=   q^0(x) - \frac{q^{1}(x) + q^{-1}(x)}{2} &\equiv 
b^q_1(x)\,,
\label{eq:forward}
\end{alignat}  
where $q_{\rightarrow [\leftarrow]}^\Lambda (x)$ represents the number 
density of a $\{$anti$\}$ quark with momentum fraction $\{x<0\}$ $x>0$ and 
positive $[$negative$]$ helicity in a rapidly moving deuteron target with 
longitudinal spin projection $\Lambda$. The `unpolarized' (polarization 
averaged) quark densities $q^\Lambda$ are defined as $\smash{ 
q^{\Lambda}(x) = q_{\rightarrow}^\Lambda(x) + q_\leftarrow^\Lambda(x) }$. 
While the probabilistic interpretation of polarization-averaged and
polarization-difference structure functions $f_1(x)$ and $g_1(x)$ in terms 
of quark densities is similar to that in the spin-1/2 case, the tensor 
structure function $b_1(x)$ does not exist for spin-1/2 targets. It has 
been measured in DIS on a polarized spin-1 target~\cite{hermes:b1}. Both 
$H_3$ and $H_5$ are associated with the 5$\%$ $D$-wave component of the 
deuteron wave function in terms of nucleons~\cite{Lacombe:1981}. ${H}_3$ 
is related to isoscalar currents and probes the binding forces in the 
deuteron, and ${H}_5$ involves a tensor 
term~\cite{Berger:2001zb,theor_deu}, the analog of which has no 
relationship to any local current due to Lorentz invariance.

\subsection{ Deeply virtual Compton scattering amplitudes }
\label{subsec:DVCS-amplitude}
For a target of atomic mass number $A$, the cross section for the hard 
exclusive leptoproduction of real photons is given by 
\cite {theor_deu,DVCS0}
\begin{equation} \label {total_gamma_xsect}
\frac{\rd \sigma}{\rd x_A \, \rd Q^2 \, \rd |t| \, \rd \phi} =
\frac {x_A \, e^6} {32 \, (2 \pi)^4 \, Q^4}
\frac {\left| \amp2 \right|^2} {\sqrt{1 + \varepsilon^2}} \,,
\end{equation}
where $x_A \equiv Q^2/(2M_A\nu)$ is the nuclear Bjorken $x_B$, where $M_A$ 
is the target mass and $\nu \equiv p \cdot q/M_A$, $\varepsilon\equiv 2 
x_A M_A/\sqrt{Q^2}$, and $\left| \amp2 \right|$ is the total reaction 
amplitude. 

As the final states of the DVCS and BH processes are indistinguishable, 
the cross section contains the square of the coherent sum of their 
amplitudes:
\begin{equation} \label {eqn:tau}
\left| \amp2 \right|^2 =  \left| \amp2_{\rmBH} +  
\amp2_{\rmDVCS} \right|^2 =
\left| \amp2_{\rmBH} \right|^2 +
\left| \amp2_{\rmDVCS} \right|^2 + \underbrace{
\amp2_{\rmDVCS} \, \amp2_{\rmBH}^* + \amp2_{\rmDVCS}^* \, 
\amp2_{\rmBH}}_{\mathrm{I}}\,.
\end{equation}
Here, $\rm I$ denotes the BH-DVCS interference term. The BH amplitude is 
calculable to leading order in Quantum Electrodynamics (QED) using nuclear 
form factors measured in elastic scattering.

The interference term $\rm I$ in Eq.~\ref{eqn:tau} provides separate 
experimental access to the real and imaginary parts of the DVCS amplitude 
through measurements of various cross-section asymmetries as functions of 
the azimuthal angle $\phi$~\cite{DVCS0}. Each of the three terms of 
Eq.~\ref{eqn:tau} can be written as a Fourier series in 
$\phi$~\cite{DVCS2}, which in the case of an unpolarized target reads
\begin{eqnarray} 
&&|\amp2^{}_{\rmBH}|^2 =
\frac{K_{\rmBH}}{{\cal P}_1(\phi){\cal P}_2(\phi)} \times
\sum_{n=0}^2 c_{n}^{\rmBH} \cos(n\phi) \label{eq:moments-BH}\, , \\
&&|\amp2^{}_{\rmDVCS}|^2 = K_{\rmDVCS} \times \Big\{ c_{0}^{\rmDVCS} + 
\sum_{n=1}^2 c_{n}^{\rmDVCS} \cos(n\phi) +
\lambda s_{1}^{\rmDVCS} \sin \phi \label{eq:moments-DVCS} \Big\}\, , \\
&&\mathrm{I} = -\frac{K_{\mathrm{I}} e_\ell}{{\cal P}_1(\phi){\cal P}_2(\phi)} \times \Big\{ 
c_{0}^{\rmI} + \sum_{n=1}^3 c_{n}^{\rmI} \cos(n\phi) +
\lambda \sum_{n=1}^2 s_{n}^{\rmI} \sin(n\phi) \Big\}\, . \label{eq:moments-I}
\end{eqnarray}
Here, $K_{\rmBH}$, $K_{\rmDVCS}$, and $K_{\mathrm{I}}$ are kinematic factors, 
$e_\ell$ denotes the lepton beam charge in units of the elementary charge, 
and $\lambda$ the helicity of the longitudinally polarized lepton beam. 
The squared BH and interference terms have an additional $\cos\phi$ dependence in the 
denominator due to the lepton propagators ${\cal P}_1(\phi)$ and ${\cal 
P}_2(\phi)$ in the BH process \cite{DVCS2,DVCS0}.

The Fourier coefficients $c_n^{\rm I}$ and $s_n^{\rm I}$ in 
Eq.~\ref{eq:moments-I} can be expressed as linear combinations of Compton 
Form Factors ${\mathcal{F}}(\xi,t)$ (CFFs)~\cite{theor_deu}, which in turn 
are convolutions of the corresponding GPDs $F^q(x,\xi,t)$ with the hard 
scattering coefficient functions ${\cal C}_q^\mp$~\cite{Bel98,Ji98a,Man98b}: 
\begin{align}
\label{eq:CFFH}
{\mathcal{F}}(\xi,t) = \sum_q \int_{-1}^1 \de x\ {\cal C}_q^\mp (\xi,x) 
F^q(x,\xi,t),
\end{align}
where the $-\{+\}$ sign applies to $F^q={H^q_1,\dots,H^q_5}$ 
$\left\{\widetilde{H}^q_1,\dots,\widetilde{H}^q_4\right\}$ in the case of 
a spin-1 target. The real and imaginary parts of the CFFs have different 
relationships to the flavor sum over the respective quark GPDs. To leading 
order in $\alpha_s$,
\begin{equation}
\label{eq:CFF1a}
 \Im\mbox{m}\left\{ \mathcal{F}(\xi,t) \right\} =
  - \pi \sum_q e_q^2\left( {F^q}(\xi,\xi,t) \mp 
{F^q}(-\xi,\xi,t)\right)\,.
\end{equation}
Hence measurements of cross-section asymmetries with respect to the beam 
helicity directly determine combinations of GPDs along the lines 
$x=\pm\xi$. In contrast, the real parts of the CFFs involve the full 
interval in $x$ and constrain the $x$ dependence of GPDs through 
convolutions:
\begin{equation}
\label{eq:CFF2a}
\Re\mbox{e}\left\{ \mathcal{F}(\xi,t) \right\} =
\sum_q e_q^2
\left[ P \int_{-1}^1 \de x \ {F^q}(x,\xi,t)
\left( \frac{1}{x-\xi} \pm \frac{1}{x+\xi} \right) \right],
\end{equation}
to leading order in $\alpha_s$. Here, $P$ denotes Cauchy's principal value. 
Since the $x$ dependence of GPDs is thereby only weakly constrained, 
experimental asymmetries in beam charge must be compared to the predictions
of various GPD models.

At leading twist (twist-2), the coefficients $c_1^{\rmI}$ and $s_1^{\rmI}$ 
are related to the same combination of GPDs.  This is also true for the 
kinematically suppressed coefficient $c_0^{\rmI}\propto 
-\frac{\sqrt{-t}}{Q}c_1^{\rmI}$. The coefficients  $c_0^{\rmI}$ and 
$c_1^{\rmI}$ are sensitive to the 
`D-term'~\cite{Polyakov:1999gs,theor_bsa1}, which contributes only in the 
`ERBL' region $-\xi<x<\xi$ where quark GPDs have the characteristics of 
distribution amplitudes for the creation of a quark-antiquark pair. It 
does not contribute in the complementary `DGLAP' region $|x|>\xi$, where 
quark GPDs describe the emission and reabsorption of an (anti-)quark in 
the infinite momentum frame, thereby having properties analogous to the 
familiar (anti-)quark distribution functions. The D-term provides a 
convenient means of representing this profound difference in GPD 
properties between the two regions, while, e.g., the absorption of this 
contribution into the double distributions~\cite{Mul94,Rad97} would 
require the introduction of terms with unnatural divergence, having a 
severity beyond representation by delta functions or their derivatives. 
In addition to $c_1^{\rmI}$, $s_1^{\rmI}$, and $c_0^{\rmI}$, the only other 
Fourier coefficient related to only twist-2 quark GPDs is $c_0^{\rmDVCS}$. 
The coefficients $c_1^{\rmDVCS}$, $s_1^{\rmDVCS}$, $c_2^{\rmI}$, and 
$s_2^{\rmI}$ appear at the twist-3 level, while $c_2^{\rmDVCS}$ and 
$c_3^{\rmI}$ arise from the gluonic transversity 
operator~\cite{Hoodbhoy,BeliMul,Diehl:2001} at twist-2 level. The highest 
harmonics of the interference and squared DVCS terms may also receive a 
twist-4 contribution~\cite{KivelMan}.

\subsection{Azimuthal cross section asymmetries}
\label{subsec:AzimuthalCrossSectionAsymmetries}
The beam-helicity asymmetries for a longitudinally (L) polarized lepton 
beam and an unpolarized (U) target, based on the difference and sum of 
yields for the two beam charges, respectively, are defined as
\begin{equation}
\label{eq:bsa1i}
\CalALUI(\phi)  \equiv  \frac
{ [ \rd \sigma^{+\rightarrow}(\phi) - 
\rd \sigma^{+\leftarrow}(\phi) ] - 
[ \rd \sigma^{-\rightarrow}(\phi) - 
\rd \sigma^{-\leftarrow}(\phi) ]}
{ [ \rd \sigma^{+\rightarrow}(\phi) +
\rd \sigma^{+\leftarrow}(\phi) ] + 
[ \rd \sigma^{-\rightarrow}(\phi) + 
\rd \sigma^{-\leftarrow}(\phi) ]}\, ,
\end{equation}
\begin{equation}
\label{eq:bsa1dvcs}
\CalALUDVCS(\phi)  \equiv  \frac
{ [ \rd \sigma^{+\rightarrow}(\phi) -
\rd \sigma^{+\leftarrow}(\phi) ] +
[ \rd \sigma^{-\rightarrow}(\phi) -
\rd \sigma^{-\leftarrow}(\phi) ]}
{ [ \rd \sigma^{+\rightarrow}(\phi) +
\rd \sigma^{+\leftarrow}(\phi) ] +
[ \rd \sigma^{-\rightarrow}(\phi) +
\rd \sigma^{-\leftarrow}(\phi) ]}\, ,
\end{equation}
where $\rightarrow$ ($\leftarrow$) denotes positive (negative) beam 
helicity and the superscript + ($-$) corresponds to  positron (electron) 
beam. These definitions serve to separate the $\sin (n\phi)$ terms in 
Eqs.~\ref{eq:moments-DVCS} and \ref{eq:moments-I}. Similarly, the 
beam-charge asymmetry (BCA) for an {\em unpolarized} beam scattering from 
this target is defined as
\begin{align}
\label{eq:bca1}
\CalACU(\phi) & \equiv  \frac
  {\rd \sigma^+(\phi) - \rd \sigma^-(\phi)}
  { \rd\sigma^+(\phi) + \rd \sigma^-(\phi)} \\ \nonumber
& =  \frac 
{ [ \rd \sigma^{+\rightarrow}(\phi) +
\rd \sigma^{+\leftarrow}(\phi) ] -
[ \rd \sigma^{-\rightarrow}(\phi) +
\rd \sigma^{-\leftarrow}(\phi) ]}
{ [ \rd \sigma^{+\rightarrow}(\phi) +
\rd \sigma^{+\leftarrow}(\phi) ] +
[ \rd \sigma^{-\rightarrow}(\phi) +
\rd \sigma^{-\leftarrow}(\phi) ]}\, .
\end{align}
In terms of the Fourier coefficients of 
Eqs.~\ref{eq:moments-BH}--\ref{eq:moments-I} these equations read as
\begin{align}
\CalALUI(\phi) & = & \frac{
- \frac{K_{\mathrm{I}}}{{\cal P}_1(\phi){\cal P}_2(\phi)} \sum_{n=1}^2 
s_{n}^{\rmI} \sin(n\phi)}{
\frac{K_{\rmBH}}{{\cal P}_1(\phi){\cal P}_2(\phi)} 
\sum_{n=0}^2 c_{n}^{\rmBH} \cos(n\phi) 
+ K_{\rmDVCS} \sum_{n=0}^2 c_{n}^{\rmDVCS} \cos(n\phi)}\, ,
\label{eq:ALUIFourier} \\
\CalALUDVCS(\phi) & = & \frac{
K_{\rmDVCS} \, s_{1}^{\rmDVCS} \sin \phi}{
\frac{K_{\rmBH}}{{\cal P}_1(\phi){\cal P}_2(\phi)}
\sum_{n=0}^2 c_{n}^{\rmBH} \cos(n\phi)
+ K_{\rmDVCS} \sum_{n=0}^2 c_{n}^{\rmDVCS} \cos(n\phi)}\, ,
\label{eq:ALUDVCSFourier} \\
\CalACU(\phi) & = & \frac{
- \frac{K_{\mathrm{I}}}{{\cal P}_1(\phi){\cal P}_2(\phi)} \sum_{n=0}^3 c_{n}^{\rmI} 
\cos(n\phi)}{
\frac{K_{\rmBH}}{{\cal P}_1(\phi){\cal P}_2(\phi)}
\sum_{n=0}^2 c_{n}^{\rmBH} \cos(n\phi) 
+ K_{\rmDVCS} \sum_{n=0}^2 c_{n}^{\rmDVCS} \cos(n\phi)}\, . 
\label{eq:ACUFourier}
\end{align}
At leading twist (twist-2, twist-3, and twist-2, respectively in the preceding three equations), and 
neglecting gluonic terms, they reduce to
\begin{equation}
\CalALUI(\phi) \simeq \frac{
- \frac{K_{\mathrm{I}}}{{\cal P}_1(\phi){\cal P}_2(\phi)} s_{1}^{\rmI} \sin \phi}{
\frac{K_{\rmBH}}{{\cal P}_1(\phi){\cal P}_2(\phi)} 
\sum_{n=0}^2 c_{n}^{\rmBH} \cos(n\phi) + K_{\rmDVCS} \, 
c_{0}^{\rmDVCS}}\, ,
\label{eq:ALUIFourierLO}
\end{equation}
\begin{equation}
\CalALUDVCS(\phi) \simeq \frac{
K_{\rmDVCS} \, s_{1}^{\rmDVCS} \sin \phi}{
\frac{K_{\rmBH}}{{\cal P}_1(\phi){\cal P}_2(\phi)} 
\sum_{n=0}^2 c_{n}^{\rmBH} \cos(n\phi) + K_{\rmDVCS} \, 
c_{0}^{\rmDVCS}}\, ,
\label{eq:ALUDVCSFourierLO}
\end{equation}
\begin{equation}
\CalACU(\phi) \simeq \frac{
- \frac{K_{\mathrm{I}}}{{\cal P}_1(\phi){\cal P}_2(\phi)} (c_{0}^{\rmI} + c_{1}^{\rmI} 
\cos \phi)}{
\frac{K_{\rmBH}}{{\cal P}_1(\phi){\cal P}_2(\phi)}
\sum_{n=0}^2 c_{n}^{\rmBH} \cos(n\phi)
+ K_{\rmDVCS} \, c_{0}^{\rmDVCS}} \, . \label{eq:ACUFourierLO}
\end{equation}

To the extent that the DVCS contributions to the common denominator
can be neglected at HERMES kinematics, the lepton propagators ${\cal 
P}_1(\phi) $ and $ {\cal P}_2(\phi)$ cancel in 
Eqs.~\ref{eq:ALUIFourier},~\ref{eq:ACUFourier},~\ref{eq:ALUIFourierLO}, 
and \ref{eq:ACUFourierLO}. However, this approximation is not invoked
in the following because it would be subject to substantial model 
uncertainty.

\subsection{From Compton form factors to asymmetries}
Measured asymmetries are used to constrain GPD models by direct comparison 
of the data with model predictions. However, it is instructive to consider 
certain approximations relating CFFs and thereby GPDs to observed 
asymmetries. (These approximations are not needed in the comparison of GPD 
model predictions with measured asymmetries.)

For an unpolarized nucleon target, the photon-helicity-conserving 
amplitude $\widetilde{\cal M}^{1,1}$ is given at leading twist by a linear 
combination of the CFFs ${\cal H}$, $\widetilde{\cal H}$ and ${\cal E}$, 
together with the Dirac and Pauli form factors $F_1$ and 
$F_2$~\cite{DVCS2}:
\begin{equation} 
\widetilde{\cal M}^{1,1} = F_1 \, {\cal H} + \frac{x_N}{2-x_N}(F_1
+F_2) \, \widetilde{\cal H} - \frac{t}{4 M_N^2} F_2 \, {\cal E}\,,
\label{eq:m11} 
\end{equation}
where $x_N$ is the Bjorken variable for the nucleon and $M_N$ is the 
nucleon mass. At small values of $x_N$ and $-t$, $\widetilde{\cal M}^{1,1} 
\simeq F_1 \, {\cal H}$ for the proton. For the neutron, the term 
containing the CFF ${\cal E}$ in Eq.~\ref{eq:m11} becomes substantial at 
large $-t$ due to the relative magnitudes of the form factors $F_1$ and 
$F_2$ for the neutron. The leading Fourier coefficients of the 
interference term can be approximated as $s_1^{\rmI} \propto \Im\mbox{m} 
\widetilde{\cal M}^{1,1}$ and $c_1^{\rmI}  \propto \Re\mbox{e} 
\widetilde{\cal M}^{1,1}$. To leading order in 1/Q and in HERMES kinematic conditions, 
\begin{align}
\CalALUI(\phi) & \propto - 
\frac{\Im\mbox{m}{\cal H}}{F_1} \sin\phi\,, \label{eq:nucl-alui} \\
\CalACU(\phi) & \propto
\frac{\Re\mbox{e}{\cal H}}{F_1} \cos\phi\,. \label{eq:nucl-acu}
\end{align}

\begin{figure}[t]   \center
\includegraphics[width=0.6\columnwidth]{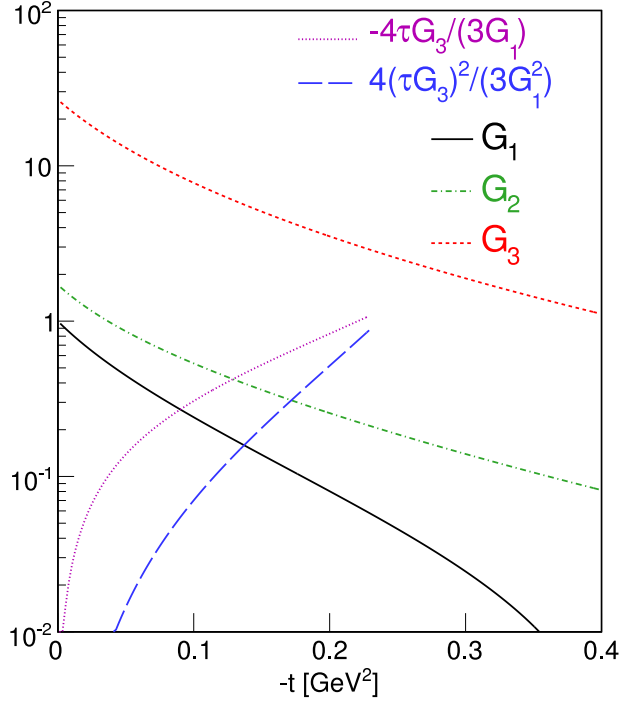}
\caption{The deuteron elastic form factors according to
Parameterization II of Ref.~\protect\cite{Deutformfac}, and the relative
contributions to the denominator of, e.g., Eq.~\ref{eq:Dtilde}
of certain terms involving $G_3$ that are not leading in $\tau$.}
\label{fig:Deutformfac}
\end{figure}

For the coherent process on the deuteron, the relationship between the 
Fourier coefficients and the GPDs is complicated. However, the 
coefficients can be expanded in powers of $x_D$, the Bjorken variable for 
the deuteron target, and $\tau= t/(4 M_{D}^2)$, where $M_D$ is the 
deuteron mass~\cite{theor_deu}. Then, to leading order in $\alpha_s$ and 
$1/Q$, $\CalALUI(\phi)$ can be expressed in terms of the imaginary part of 
the deuteron CFFs ${\cal H}_1$, ${\cal H}_3$ and ${\cal H}_5$ and the 
deuteron elastic form factors~\cite{Deutformfac} $G_1$ and $G_3$ (see 
Fig.~\ref{fig:Deutformfac}). The quantity $|\tau|$ is typically about 
0.003 in the range of small $-t$ where the coherent process is significant,
extending up to values of $\tau$ only as large as 0.01. However, as shown in 
Fig.~\ref{fig:Deutformfac}, the magnitude of $G_3$ exceeds that of $G_1$ 
by more than one order of magnitude. Hence certain terms leading in $\tau$ 
(but not $x_D$) are retained. Defining
\begin{equation}
\widetilde{\cal D}_{\rm U}^{1,1} \equiv \frac{ 3 G_1 {\cal H}_1
 - 2 \tau [G_1 {\cal H}_3 + G_3({\cal H}_1 - \frac{1}{3} {\cal H}_5)]
 + 4\tau^2 G_3 {\cal H}_3 }
 {3 G_1^2 - 4\tau G_1 G_3 + 4\tau^2 G_3^2 }, \label{eq:Dtilde}
\end{equation} 
the kinematic expansion yields 
\begin{equation}
\label{eq:App-ALU} 
\CalALUI(\phi) \simeq  -
\frac{x_{D} (2-y)
\sqrt{\frac{-t}{Q^2}(1-y)}}{2-2y + y^2}
\Im\mbox{m} \widetilde{\cal D}_{\rm U}^{1,1} \sin \phi \, ,
\end{equation}
where $y \equiv p \cdot q/(p \cdot k)$.

The relative contributions of those terms are also shown in 
Fig.~\ref{fig:Deutformfac}; they are less than 10\% at 
$-t<0.03\rm\,GeV^2$. When these terms are neglected, Eq.~\ref{eq:App-ALU} 
becomes
\begin{equation}
\CalALUI(\phi) \simeq -
\frac{x_{D} (2-y)
\sqrt{\frac{-t}{Q^2}(1-y)}}{2-2y + y^2}
\frac{\Im\mbox{m}{\cal H}_1}{G_1} \sin \phi \, .  
\label{eq:DUtilde-short}
\end{equation}

The deuteron $\CalACU(\phi)$ is related to  the real part of the same 
linear combination of CFFs appearing in the deuteron 
$\CalALUI(\phi)$:
\begin{eqnarray}
\label{App-AC}
\CalACU(\phi) &\simeq&
-\frac{x_{D} \sqrt{\frac{-t}{Q^2}(1-y)}}{y}
\Re\mbox{e}  \widetilde{\cal D}_{\rm U}^{1,1} \cos \phi  \\
 &\simeq& -\frac{x_{D} \sqrt{\frac{-t}{Q^2}(1-y)}}{y}
\frac{\Re\mbox{e} {\cal H}_1}{G_1} \cos \phi \, .
\label{eq:App-AC-short}
\end{eqnarray}

For the coherent process on the deuteron, the leading term in the 
expansion of coefficients $s_1^{\rmI}$ and $c_1^{\rmI}$ lead respectively 
to Eqs.~\ref{eq:DUtilde-short} and \ref{eq:App-AC-short} which are 
analogous to Eqs.~\ref{eq:nucl-alui} and \ref{eq:nucl-acu} for 
scattering on the nucleon.

\section{The HERMES experiment}
\label{sec:experiment}
A detailed description of the HERMES experiment can be found in 
Ref.~\cite{hermes:spectrometer}. A  longitudinally polarized positron or 
electron beam  of 27.6~GeV energy was scattered from an unpolarized 
deuterium gas target internal to  the HERA lepton storage ring at DESY. 
The lepton beam was transversely polarized via the asymmetry in the 
emission of synchrotron radiation (Sokolov-Ternov effect)~\cite{Sokolov+:1964} in the arcs of the 
HERA storage ring. The transverse beam polarization was transformed  
locally into longitudinal polarization by a pair of spin rotators located 
before and after the experiment~\cite{Buon}. The helicity of the beam was 
typically reversed approximately every two months.

The beam polarization was continuously monitored by two Compton 
backscattering polarimeters~\cite{TPOL:1994,LPOL:2002}. The average values 
of the beam polarization for various running periods are given in 
Table~\ref{tb:table1}; the average fractional systematic uncertainty was 
2.4$\%$. The scattered leptons and produced particles were detected in the 
polar angle range $0.04$~rad~$< \theta < 0.22$~rad. The lepton trigger 
required a coincidence of signals from scintillator hodoscope planes and 
the local deposition of a minimum energy of 3.5 GeV in the electromagnetic 
calorimeter. Lepton identification was accomplished using the
transition-radiation detector, the preshower scintillator counter, and the 
electromagnetic calorimeter. The average lepton identification efficiency 
was at least 98$\%$ with hadron contamination that was less than 1\%. 
Photons were identified by the detection of energy deposited in the 
calorimeter and preshower counter with no associated charged-particle 
track.

\section{Event selection and yield distributions}
\label{sec:event_yield}
\begin{table}[t] \center
\caption{The beam charge and polarization as well as the integrated 
luminosity in pb$^{-1}$ of the data sets used for the extraction of the 
various asymmetries on the unpolarized deuterium target.}
\begin{tabular}{ccrrrr}
\noalign{\smallskip}
\hline
& \hspace{1.0cm}Beam &  \multicolumn{2}{c}{\hspace{1.0cm}Beam} 
&\multicolumn{2}{c}{\hspace{1.0cm}Luminosity} \\
Year & \hspace{1.0cm}Charge & 
\multicolumn{2}{c}{\hspace{1.0cm}Polarization} & 
\multicolumn{2}{c}{\hspace{1.0cm}[pb$^{-1}$]} \\
&  & \hspace{1.0cm}$\lambda = -1$ & $\lambda = +1$ & 
\hspace{1.0cm}$\lambda = -1$ & $\lambda = +1$ \\
\hline
1996 & \hspace{1.0cm}$e^+$ & & $\, \, 0.516$ & & 43.9 \\
1997 & \hspace{1.0cm}$e^+$ & $- \, 0.511$ & & 53.1 & \\
1998 & \hspace{1.0cm}$e^-$ & $- \, 0.307$ & & 24.1 & \\
1999 & \hspace{1.0cm}$e^+$ & $- \, 0.552$ & $\, 0.418$ & 0.9 & 5.1 \\
2000 & \hspace{1.0cm}$e^+$ & $- \, 0.584$ & $\, 0.552$ & 29.7 & 9.0 \\
2005 & \hspace{1.0cm}$e^-$ & $- \, 0.355$ & $\, 0.377$ & 66.3 & 65.7 \\
\hline
Sum& & & & $174.1$ & $123.7$ \\
\hline
\end{tabular}
\label{tb:table1}
\end{table}
\begin{figure}[t]   \center
\includegraphics[width=0.68\columnwidth]{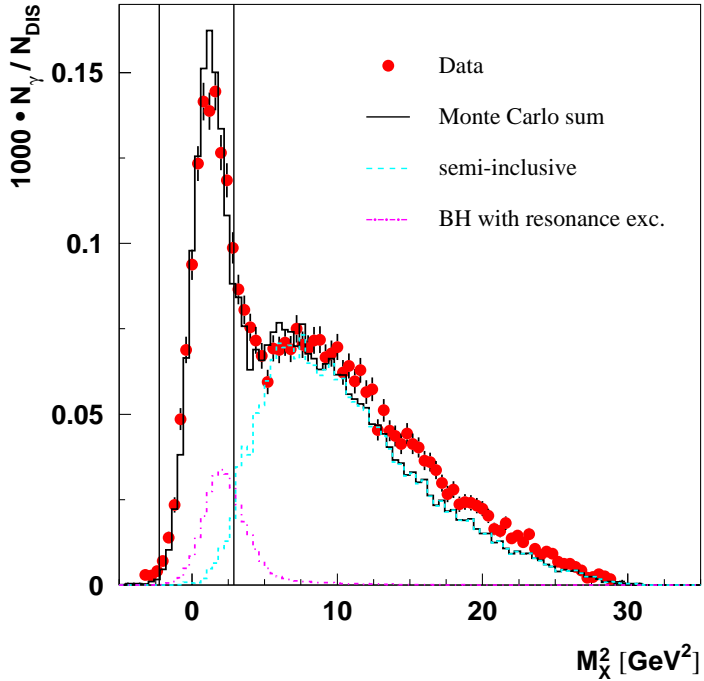}
\caption{The measured distribution (points) of electroproduced real-photon 
events versus the squared missing mass $M_X^2$. The solid curve represents 
a Monte Carlo simulation including coherent and incoherent BH and DVCS 
processes, the BH processes with the excitation of resonant final states 
(represented separately by the dashed-dotted curve), and the 
semi-inclusive background (dashed curve). The simulations and data are 
both normalized to the number of DIS events. The region between the two 
vertical lines indicates the selected exclusive events.}
\label{figure1.1}
\end{figure}
The data sets used in the extraction of the various asymmetries reported 
here are given in Table~\ref{tb:table1}. In this analysis, it was required 
that events  contained exactly one charged-particle track consistent with 
being the scattered beam lepton, and a single cluster in the calorimeter 
with an energy deposit $E_{\gamma} > 5.0\rm\,GeV$ and with no associated 
charged track. The following requirements were imposed on the event 
kinematics: $1{\rm\,GeV^2} < Q^2 <10\rm\,GeV^2 $, $W_N^2 >9\rm\,GeV^2$, 
$\nu<22$\,GeV and $ 0.03 < x_N <0.35 $, where $W_N^2 = M_N^2 + 2M_N\nu - 
Q^2$, $x_N = Q^2/(2M_{N} \nu)$, and $\nu \equiv p \cdot q/M_N$. The 
nucleonic (proton) mass $M_N$ was used in all kinematic constraints on event 
selection even at small values of $-t$, where coherent reactions on the 
deuteron are dominant, because the experiment did not distinguish between 
coherent and incoherent scattering and the latter dominates over most of 
the kinematic range. Monte Carlo studies have shown that this choice has 
little effect on the extracted asymmetries~\cite{Bernie}. In order to 
reduce background from the decay of neutral mesons, the angle between the 
laboratory 3-momenta of the real and virtual photons was limited to 
$\theta_{\gamma^* \gamma} < 45$\,mrad.  The minimum angle requirement 
$\theta_{\gamma^* \gamma} > 5$\,mrad was chosen according to Monte Carlo 
studies to be compatible with the effects of instrumental resolution in 
determination of $\phi$.

`Exclusive' single-photon events were selected by requiring the squared 
missing mass $M_X^2$ to be close to the squared nucleon mass $M_N^2$, 
where $M_X^2$ is defined as $M_X^2 = (q + P_N - q^\prime)^2$ with $P_N = 
(M_N,0,0,0)$. Due to the finite resolution of the spectrometer and the 
calorimeter, $M_X^2$ may be negative. In Fig.~\ref{figure1.1}, the 
squared missing mass distribution of the selected events is compared with 
the predictions of Monte Carlo simulations of processes that contribute to 
both signal and background. One of the simulations uses an exclusive-photon 
generator for the BH and DVCS processes, including coherent and 
incoherent reactions as well as the excitation of resonant final states (a 
category known as associated production). The DVCS simulation for 
incoherent reactions on the proton is based on Ref.~\cite{GPD:PROTON}, 
while that for coherent reactions on the deuteron is based on the model 
from Ref.~\cite{theor_deu}. Most of the background in the vicinity of the 
exclusive peak comes from the decay of neutral pions. The dominant source 
of neutral pions is semi-inclusive DIS, $\gamma^{\ast}N \to \pi^{0}X \to 
\gamma \gamma X$, which is simulated using the {\sc Lepto} event 
generator~\cite{LEPTO} with a set of {\sc Jetset}~\cite{Jetset} 
fragmentation parameters tuned for HERMES kinematic 
conditions~\cite{ourtune}. In this simulation, the photon originates 
mainly from decay of $\pi^0$s from DIS fragmentation. Incoherent exclusive 
$\pi^0$ production, $\gamma^{\ast}N \to \pi^{0}N$, was simulated using an 
exclusive Monte Carlo event generator based on the GPD models of 
Ref.~\cite{Vanderhaeghen:1999xj} and was found to be 
negligible~\cite{Bernie,Zhenyu}. HERMES data support this 
estimate~\cite{Arne}. The Monte Carlo yield exceeds the data by 
approximately 2$\%$ in the exclusive region. This may be due to the 
contribution of the DVCS process in the simulation of both coherent and 
incoherent processes, which is highly model-dependent and can vary between 
10$\%$ and 25$\%$~\cite{Zhenyu} for the incoherent processes. On the other 
hand, radiative effects not included in the simulation would move events 
from the peak to the continuum~\cite{Vanderhaeghen:2000}.

Events were selected in the `exclusive region', defined as 
$-(1.5)^2{\rm\,GeV}^2 < M_X^2 <(1.7)^2{\rm\,GeV}^2$ to minimize background 
from DIS fragmentation while maintaining reasonable 
efficiency~\cite{Frank}.
\begin{figure}[!tb]   \center
\includegraphics[width=0.7\columnwidth]{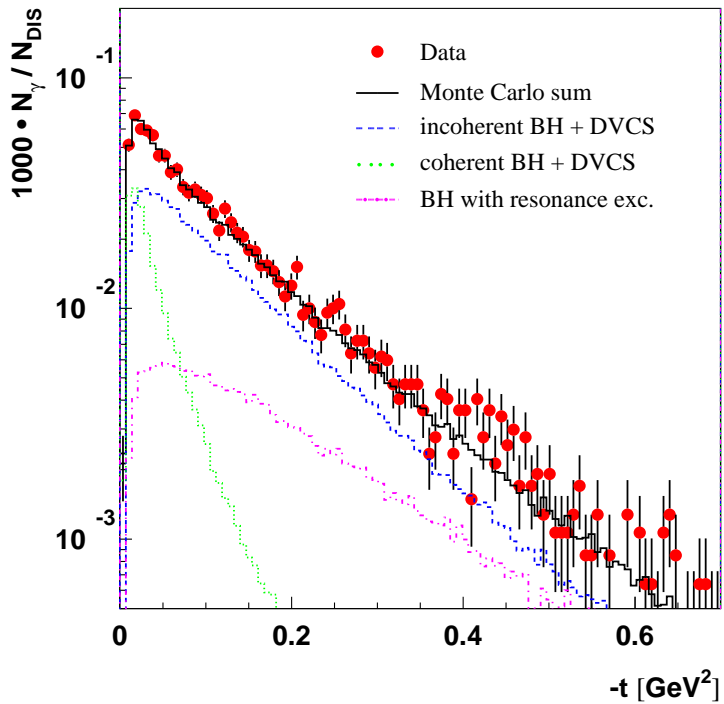}
\caption{Distribution in $-t$ of events selected in the exclusive 
region of $M_X^2$. The points represent experimental data while the
continuous curve represents the simulation of real-photon production for 
all exclusive final states including resonances. Background from $\pi^0$ 
decay is not included. The dotted and dashed curves represent the BH plus 
DVCS contributions of the coherent and incoherent elastic process, 
respectively. The dash-dotted curve shows the resonant BH contributions. 
The simulations and data are both normalized to the number of DIS events.}
\label{figure1.2}
\end{figure}

As the recoiling target nucleon or nucleus was undetected, the Mandelstam 
variable $t$ must be reconstructed from the measured kinematics of the 
scattered lepton and the detected photon. The resolution in the photon 
energy from the calorimeter is inadequate for a precise determination of 
$t$. Hence for events selected in the exclusive region in $M_X^2$, the 
final state is assumed to be exclusive, leaving the target intact, 
thereby allowing $t$ to be reconstructed with improved resolution using 
only the photon direction and the lepton kinematics~\cite{hermes_bca_2006}: 
\begin{equation}
t = \frac{-Q^2 - 2 \, \nu \, (\nu - \sqrt{\nu^2 + Q^2} \, 
\cos\theta_{\gamma^* \gamma })}
{1 + \frac{1}{M_N} \, (\nu - \sqrt{\nu^2 + Q^2} \, \cos\theta_{\gamma^* 
\gamma}  \, 
)}.\label{tc}
\end{equation}
The further restriction $-t < 0.7\rm\,GeV^2$ is imposed in the selection 
of exclusive events in order to reduce background from the decay of neutral mesons.

The $t$ distribution of events for the deuterium target is shown in 
Fig.~\ref{figure1.2} and compared with the Monte Carlo simulations 
discussed above. The simulated contributions of coherent and incoherent 
processes on the deuteron are also shown separately. Coherent scattering 
on the deuteron occurs preferentially at small values of $-t$. The Monte 
Carlo simulation shows that requiring $-t < 0.06\rm\,GeV^2$ enhances the 
mean fractional contribution of the coherent process from 20$\%$ to 40$\%$ 
in the HERMES spectrometer acceptance. Requiring $-t < 0.01\rm\,GeV^2$ can 
further enhance the coherent contribution to 66$\%$, but only at the cost 
of a rapidly decreasing yield. In Sections~\ref{subsec:result_unpol} and 
~\ref{subsec:comparison_p_d}, the first two $-t$ bins covering the range 
$0.00-0.06\rm\,GeV^2$ will provide a measure of coherent effects; in 
Section~\ref{subsec:estimates_coh}, an attempt is made to isolate the 
coherent contribution.

\section{Analysis of the data}
\subsection{Extraction of azimuthal asymmetry amplitudes}
The distribution of the expectation value of the yield for scattering a 
polarized lepton beam from an unpolarized deuterium target is given by
\begin{eqnarray}
\langle\intN\rangle(P_\ell,e_\ell,\phi) & = & \Lumi(P_\ell,e_\ell)\,
\eta(e_\ell,\phi)\,\CUU(\phi) \nonumber \\
&\times& 
\bigl[1+P_\ell\CalALUDVCS(\phi)+e_\ell\CalACU(\phi)+e_\ell P_\ell\CalALUI
(\phi)\bigr].\label{eq:A1}
\end{eqnarray}
Here, $\Lumi$ denotes the integrated luminosity, $P_\ell$ the 
longitudinal beam polarization, $\eta$ the detection efficiency, and 
$\CUU(\phi)$ the cross section for an unpolarized target averaged over 
both beam charges and both beam helicities, which can be expressed as 
\begin{eqnarray}
& \CUU &(\phi) = 
\frac {x_D} {32 \, (2 \pi)^4 \, Q^4}
\frac {1} {\sqrt{1 + \varepsilon^2}} \nonumber \\
&\times& \bigg\{ \frac{K_{\rmBH}}{{\cal P}_1(\phi){\cal P}_2(\phi)}
\sum_{n=0}^2 c_{n}^{\rmBH} \cos(n\phi)
+ K_{\rmDVCS} \sum_{n=0}^2 c_{n}^{\rmDVCS} \cos(n\phi) \bigg\}.
\label{eq:sigma00}
\end{eqnarray}
The asymmetries $\CalALUI(\phi)$, $\CalALUDVCS(\phi)$, and $\CalACU(\phi)$ 
are related to the Fourier coefficients appearing in 
Eqs.~\ref{eq:moments-BH}--\ref{eq:moments-I}, as illustrated by 
Eqs.~\ref{eq:ALUIFourier}--\ref{eq:ACUFourier}. In analogy to the 
expansion of the cross section in Eq. 7-9, these asymmetries are 
also expanded in terms of the same harmonics in $\phi$:
\begin{eqnarray}
\label{eq:sigma00_assi}
& & \CalALUI(\phi) \simeq \sum_{n=1}^2 \ALUI^{\sin(n\phi)}\sin(n\phi) + 
\ALUI^{\cos(0\phi)}\, , \\
\label{eq:sigma00_assdvcs}
& & \CalALUDVCS(\phi) \simeq \ALUDVCS^{\sin \phi}\sin \phi + 
\ALUDVCS^{\cos(0\phi)}\, , \\
\label{eq:sigma00_assc}
& & \CalACU(\phi) \simeq \sum_{n=0}^3 \ACU^{\cos(n\phi)}\cos(n\phi)\, ,
\end{eqnarray}
where the approximation is due to the truncation of the
in general infinite Fourier series caused by the azimuthal
dependences in the denominators of Eqs.~\ref{eq:ALUIFourier}--\ref{eq:ACUFourier}.

For each kinematic bin in $-t$, $x_B$, or $Q^2$, the sets of azimuthal 
asymmetry amplitudes $\ALUI^{\sin(n\phi)}$, $\ALUDVCS^{\sin \phi}$ and 
$\ACU^{\cos(n\phi)}$, hereafter called `asymmetry amplitudes', are 
simultaneously extracted from the observed exclusive sample using the 
method of maximum likelihood (described in detail in 
Ref.~\cite{hermes_ttsa}). Although these asymmetry amplitudes differ 
somewhat from the coefficients given in 
Eqs.~\ref{eq:moments-BH}--\ref{eq:moments-I} and 
Eqs.~\ref{eq:ALUIFourier}--\ref{eq:ACUFourier}, they are well defined and 
can be computed in various GPD models for direct comparison with the data. 
Note that in Eqs.~\ref{eq:sigma00_assi} and \ref{eq:sigma00_assdvcs}, an 
additional constant term ($n=0$) was introduced as a consistency test. 
These terms must vanish as they are parity violating. 
Removing these constant terms or also introducing additional 
harmonic terms in the fitting procedure do not influence results for other 
asymmetry amplitudes~\cite{Gordon_thesis}.

\subsection{Background corrections and systematic uncertainties}
\label{subsec:background}
In each kinematic bin, the results from the maximum likelihood fit are 
corrected for photon background arising from semi-inclusive production of 
neutral mesons, mainly pions. A corrected asymmetry amplitude is obtained 
as
\begin{equation}
A_{\rm corr} = \frac{A_{\rm raw} - f_{\rm semi} \cdot 
A_{\rm semi}}{1 - f_{\rm semi}} \, .
\label{equ:pi}
\end{equation}
Here, $A_{\rm raw}$ stands for the extracted raw asymmetry amplitude, and 
$f_{\rm semi}$ and $A_{\rm semi}$ the fractional contribution and 
corresponding asymmetry amplitude of the semi-inclusive background. This 
fraction is obtained from a Monte Carlo simulation (see 
Section~\ref{sec:event_yield}) and ranges from 1$\%$ to 11$\%$, depending 
on the kinematic conditions. As the semi-inclusive process is only very 
weakly beam-charge dependent, its asymmetry with respect to the beam 
charge or to the product of the beam charge and the beam polarization is 
assumed to be zero. The asymmetry of the semi-inclusive $\pi^0$ 
background with respect to only the longitudinal beam polarization is 
extracted from experimental data by requiring two photons to be detected 
in the calorimeter with an invariant mass between 0.10 GeV and 0.17 GeV 
and with no associated charged tracks. The restriction on the energy 
deposition in the calorimeter of the less energetic cluster is relaxed to 
1~GeV to improve the statistical precision. The fractional energy $z = 
E_{\pi}/\nu$ of the reconstructed neutral pions is required to be larger 
than 0.8. After applying the correction of Eq.~\ref{equ:pi}, the resulting 
asymmetry amplitudes are expected to originate from elastic (coherent), 
and incoherent photon production possibly including nucleon excitation.

The combined contribution to the systematic uncertainty from detector 
acceptance, smearing, finite bin width, and alignment of the detector elements with 
respect to the beam is determined from a Monte Carlo simulation using 
the GPD model described in Ref.~\cite{Guzey}. Note that a mistake has been 
found in this GPD model~\cite{Guzey1}; however, the model described 
previously reported HERMES beam-charge \cite{hermes_ttsa} and preliminary 
(single-charge) beam-helicity asymmetries well \cite{Frank1} and thus is 
considered to be adequate for systematic studies. In each bin, the 
systematic uncertainty is taken as the difference between the model 
prediction at the mean kinematic value of that bin and the respective 
amplitude extracted from the reconstructed Monte Carlo data. The dominant 
contributions to the total systematic uncertainty are those from the 
detector acceptance and finite bin width. Further sources of uncertainty 
are associated with the background correction and a relative shift of the 
$M_X^2$ spectra between the data samples from various running 
periods~\cite{hermes_ttsa}. The contributions to the systematic 
uncertainty are added in quadrature. The main contributions for asymmetry 
amplitudes of interest are given in Table~\ref{tb:table2}. Not included is 
any contribution due to additional QED vertices, as the most significant 
of these has been estimated to be negligible~\cite{Afanasev}.

\begin{table}
\caption{The main contributions to the systematic uncertainty of extracted 
asymmetry amplitudes of interest, averaged over the full kinematic range. 
Not included is a 2.4$\%$ scale uncertainty of the beam-helicity 
asymmetries due to the beam polarization measurement.}
\begin{center}
\begin{tabular}{crcc}
\noalign{\smallskip}
\hline
Amplitude & \hspace{0.5cm} $M_X^2$ shift & \hspace{0.2cm} Background corr. 
& \hspace{0.5cm} Acceptance, smearing, bin width, alignment\\
\hline
$\ACU^{\cos (0\phi)}$ & 0.001 & 0.001 & 0.014 \\
$\ACU^{\cos \phi}$ & $<$ 0.001 & 0.002 & 0.023 \\
$\ALUI^{\sin \phi}$ & $<$ 0.001 & 0.004 & 0.031 \\
$\ALUDVCS^{\sin \phi}$ & 0.002 & 0.006 & 0.003 \\
\hline
\end{tabular}
\label{tb:table2}
\end{center}
\end{table}

\section{Results}
\label{sec:results}
\subsection{Results on beam-charge and beam-helicity asymmetries 
 for an unpolarized deuterium target}
\label{subsec:result_unpol}
\begin{figure}[!tb]
\includegraphics[width=1\columnwidth]{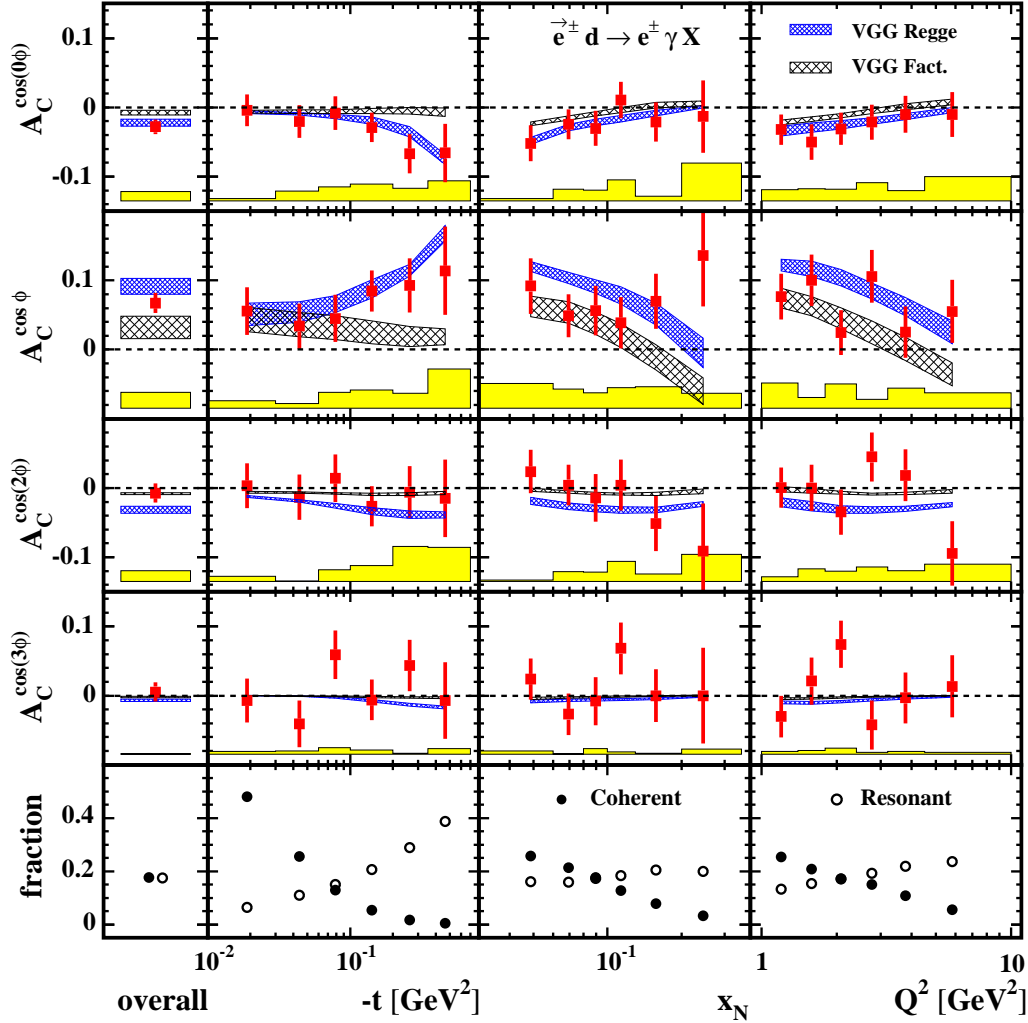}
\caption{Amplitudes of the beam-charge asymmetry, which are sensitive to 
the interference term, in bins of $-t$, $x_N$, or $Q^2$. The squares 
represent the results from the present work. The error bars (bands) 
represent the statistical (systematic) uncertainties. The finely (coarsely) hatched 
bands are theoretical calculations for incoherently combined proton and 
neutron targets, using variants of a double-distribution 
model~\cite{Vanderhaeghen:1999xj,Radyushkin:1998es,Goeke:2001tz} with 
the VGG Regge (VGG Factorized) ansatz for GPDs. The lowest panel shows 
the simulated fractions of coherent and resonant production.}
\label{fig:bca_uu}
\end{figure}

\begin{figure}[!tb]
\includegraphics[width=1\columnwidth]{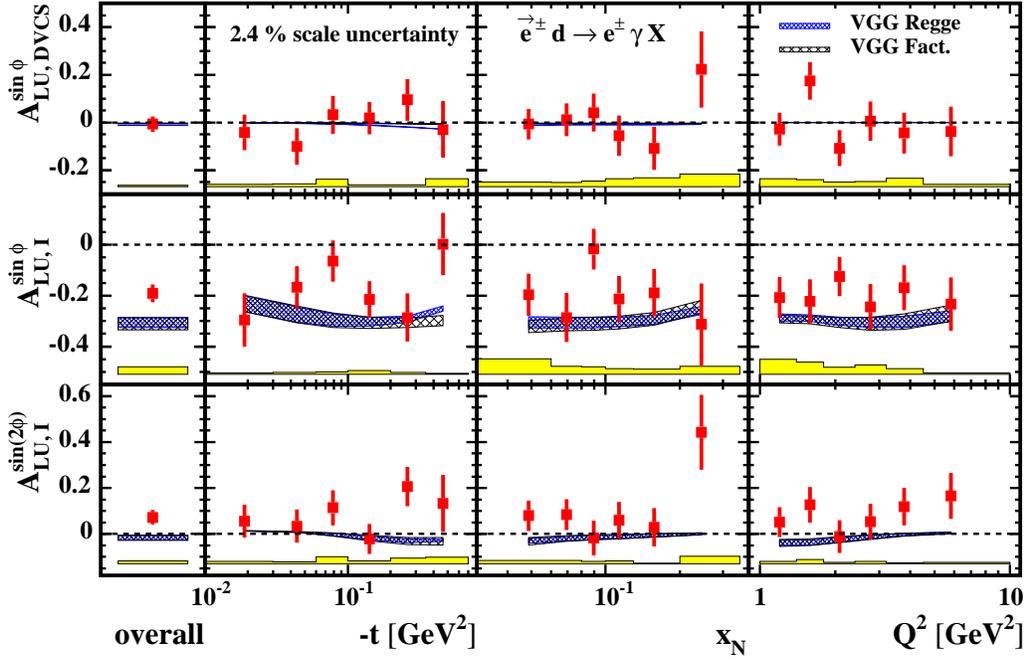}
\caption{The first row shows the $\sin \phi$ amplitude of the 
beam-helicity asymmetry that is sensitive to the squared DVCS term, in 
bins of $-t$, $x_N$, or $Q^2$. Correspondingly, the second (third) row 
shows the $\sin \phi$ ($\sin 2\phi$) amplitude of the beam-helicity 
asymmetry sensitive to the interference term. All symbols are defined as 
in Fig.~\ref{fig:bca_uu}. There is an overall 2.4$\%$ scale uncertainty 
arising from the uncertainty in the measurement of the beam polarization.}
\label{fig:bsad_uu}
\end{figure}

\begin{figure}[!tb]
\includegraphics[width=1\columnwidth]{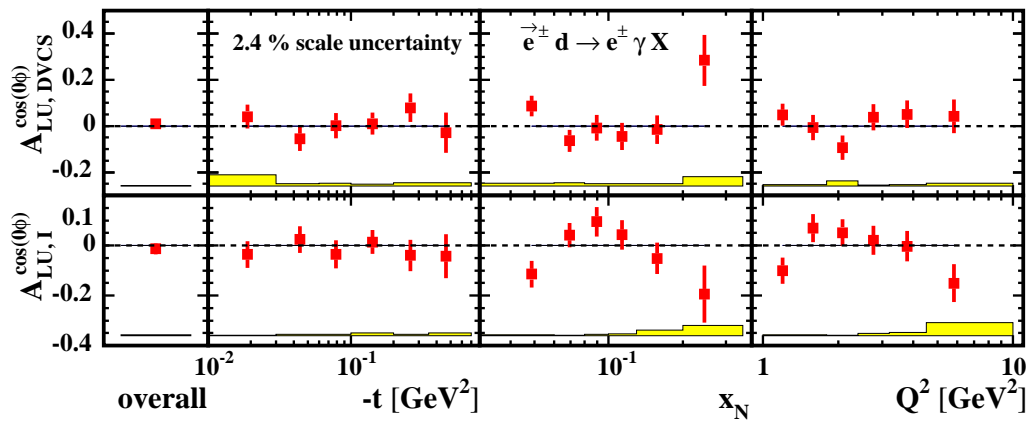}
\caption{The $\cos (0\phi)$ amplitudes (constant terms) that are included 
as a consistency test in the fit in Eqs.~\ref{eq:sigma00_assdvcs} and 
\ref{eq:sigma00_assi}. All symbols are defined as in 
Fig.~\ref{fig:bca_uu}. There is an overall 2.4$\%$ scale uncertainty 
arising from the uncertainty in the measurement of the beam 
polarization.}
\label{fig:bsai_uu}
\end{figure}

The asymmetry amplitudes are shown in 
Figs.~\ref{fig:bca_uu}--\ref{fig:bsai_uu} as a function of $-t$, $x_N$, 
or $Q^2$. While the variable $x_D$ would be the appropriate choice 
when presenting experimental results for pure coherent scattering, the 
nucleonic Bjorken variable $x_N$ is the practical choice in this case where 
incoherent scattering dominates over most of the kinematic range. The 
variables $x_N$ and $Q^2$ are strongly correlated due to the experimental 
acceptance. The `overall' results in the left columns correspond to the 
entire HERMES kinematic acceptance. Figure~\ref{fig:bca_uu} shows the 
amplitudes $\ACU^{\cos(n\phi)}$, which are related to beam charge only, 
and Fig.~\ref{fig:bsad_uu}  shows the amplitude $\ALUDVCS^{\sin \phi}$, 
which is related to beam helicity only, and the amplitudes 
$\ALUI^{\sin(n\phi)}$, which are related to both. All amplitudes are 
listed in Table~\ref{tb:table3} with the mean kinematic values of each 
bin\footnote{These results for only four bins in $-t$, $x_N$, or $Q^2$, 
i.e., a binning used in previous HERMES 
papers~\cite{hermes_ttsa,hermes_bca_2006}, are available in the Durham 
database.}.

Of special interest is the asymmetry amplitude $\ACU^{\cos \phi}$, 
which is sensitive to the GPD $H_1$ ($H$) for the coherent (incoherent) 
process in HERMES kinematic conditions (see Eqs.~\ref{eq:App-AC-short} 
and \ref{eq:nucl-acu}). The present data indicates that this 
amplitude increases with increasing $-t$. The amplitude $\ACU^{\cos 
(0\phi)}$ in Fig.~\ref{fig:bca_uu}, which is expected to relate to the 
same combination of GPDs as does $\ACU^{\cos \phi}$, shows similar 
behaviour but with opposite sign, as expected~\cite{DVCS2}. The other two 
amplitudes $\ACU^{\cos (2\phi)}$ and $\ACU^{\cos (3\phi)}$, related to 
twist-3 GPDs and the gluon transversity operator, respectively 
(see Section~\ref{subsec:DVCS-amplitude}), are consistent with zero. 

The fractional contributions to the yield from the coherent processes
and from processes with excitation of resonant final states are presented 
in the bottom row of Fig.~\ref{fig:bca_uu} (see also 
Table~\ref{tb:table4}), as obtained from the Monte Carlo simulation 
using the exclusive-photon generator mentioned in 
Section~\ref{sec:event_yield}. Note that these fractional contributions 
are subject to considerable model dependence.

Figure~\ref{fig:bsad_uu} shows amplitudes of beam-helicity asymmetries, 
with the charge-averaged case related to the squared DVCS term in the upper 
row and the charge-difference case related to the interference term in the 
other rows. The amplitude $\ALUDVCS^{\sin \phi}$, which is related to 
twist-3 GPDs, is found to be consistent with zero. Like the amplitude 
$\ACU^{\cos\phi}$, the amplitude $\ALUI^{\sin \phi}$ is also 
sensitive to the GPD $H_1$ [$H$] for the coherent [incoherent] process, 
although these two asymmetries reveal different aspects of the (real) GPD, 
selected by different convolutions with (complex) hard scattering 
amplitudes. While the amplitude $\ACU^{\cos \phi}$ is related to the real 
part of the CFF ${\cal H}_1$ [$\cal H$], the $\ALUI^{\sin \phi}$ amplitude 
is proportional to the imaginary part  and shows significant negative 
values. The amplitude $\ALUI^{\sin (2\phi)}$ appears at twist-3 level, but 
nevertheless it shows a value which is non-zero and positive by 1.7 
standard deviations of the total experimental uncertainty. 
Figure~\ref{fig:bsai_uu} shows the amplitudes that are forbidden by parity 
conservation but were included in the fit as a consistency test. They are 
consistent with zero.
\begin{table}
\caption{Results for azimuthal Fourier amplitudes of the asymmetries with 
respect to the beam charge and helicity for the exclusive sample.}
\tiny
\begin{center}
\begin{tabular}{r|ccccrrrr}
\noalign{\smallskip}
\hline\noalign{\smallskip}
\multicolumn{2}{c}{kinematic bin} 
&$\langle -t \rangle$ &$\langle x_N \rangle$ &$\langle Q^2
\rangle$ &$\ACU^{\cos \, (0\phi)}$\hspace{0.8cm}
&$\ACU^{\cos \, \phi}$\hspace{1.05cm} &$\ACU^{\cos \, 
(2\phi)}$\hspace{0.8cm} &$\ACU^{\cos \, (3\phi)}$ 
\hspace{0.7cm} \\
\multicolumn{2}{c}{} &[GeV$^2$] & &[GeV$^2$] &$\pm \rm {\delta_{stat}} \pm 
\rm {\delta_{syst}}$ &$\pm \rm {\delta_{stat}} \pm \rm {\delta_{syst}}$ 
&$\pm \rm {\delta_{stat}} \pm \rm {\delta_{syst}}$ &$\pm \rm 
{\delta_{stat}} \pm \rm {\delta_{syst}}$\\ 
\noalign{\smallskip}
\hline\noalign{\smallskip}
\multicolumn{2}{c}{overall} & 0.13 & 0.10 & 2.5 & $-0.028\pm0.010 
\pm0.014$ & $\, 0.067\pm0.015\pm0.023$ & $-0.007\pm0.014\pm0.016$ & 
$0.005\pm0.014\pm0.001$\\
\noalign{\smallskip}
\hline\noalign{\smallskip}
\multirow{6}{*}{\rotatebox{90}{\mbox{$-t$[GeV$^2$]}}} \hspace{0.01cm}
&0.00-0.03 & 0.02 & 0.07 & 1.7 & $-0.004\pm0.023\pm0.003$ 
& $0.056\pm0.035\pm0.011$ & $0.003\pm0.032\pm0.008$ & 
$-0.007\pm0.032\pm0.004$\\
&0.03-0.06 & 0.04 & 0.09 & 2.2 & $-0.020\pm0.024\pm0.014$ 
& $0.034\pm0.033\pm0.007$ & $-0.013\pm0.033\pm0.001$ & 
$-0.041\pm0.034\pm0.005$\\
&0.06-0.10 & 0.08 & 0.10 & 2.4 & $-0.008\pm0.024\pm0.020$ 
& $0.045\pm0.034\pm0.023$ & $0.014\pm0.035\pm0.017$ & 
$0.059\pm0.035\pm0.010$\\
&0.10-0.20 & 0.14 & 0.11 & 2.7 & $-0.029\pm0.021\pm0.024$ 
& $0.085\pm0.030\pm0.027$ & $-0.026\pm0.029\pm0.023$ & 
$-0.006\pm0.029\pm0.006$\\
&0.20-0.35 & 0.26 & 0.12 & 3.1 & $-0.067\pm0.028\pm0.018$
& $0.093\pm0.039\pm0.022$ & $-0.006\pm0.038\pm0.050$ & 
$0.044\pm0.037\pm0.002$\\
&0.35-0.70 & 0.46 & 0.11 & 3.5 & $-0.066\pm0.042\pm0.029$
& $0.114\pm0.064\pm0.057$ & $-0.015\pm0.056\pm0.049$ & 
$-0.007\pm0.055\pm0.008$\\
\noalign{\smallskip}
\hline\noalign{\smallskip}
\multirow{6}{*}{\rotatebox{90}{\mbox{$x_N$}}} \hspace{0.01cm}
&0.03-0.06 & 0.12 & 0.05 & 1.3 & $-0.052\pm0.026\pm0.003$ 
& $0.092\pm0.040\pm0.036$ & $0.024\pm0.031\pm0.002$ & 
$0.024\pm0.030\pm0.005$\\
&0.06-0.08 & 0.10 & 0.07 & 1.8 & $-0.024\pm0.022\pm0.017$ 
& $0.049\pm0.031\pm0.028$ & $0.004\pm0.029\pm0.014$ & 
$-0.027\pm0.030\pm0.001$\\
&0.08-0.10 & 0.11 & 0.09 & 2.3 & $-0.030\pm0.025\pm0.016$  
& $0.056\pm0.036\pm0.023$ & $-0.014\pm0.035\pm0.013$ & 
$-0.008\pm0.035\pm0.008$\\
&0.10-0.13 & 0.13 & 0.11 & 2.9 & $0.011\pm0.026\pm0.030$ 
& $0.039\pm0.037\pm0.030$ & $0.004\pm0.037\pm0.029$ & 
$0.068\pm0.037\pm0.004$\\
&0.13-0.20 & 0.17 & 0.16 & 4.0 & $-0.021\pm0.028\pm0.007$
& $0.070\pm0.040\pm0.031$ & $-0.051\pm0.040\pm0.011$ & 
$0.000\pm0.038\pm0.002$\\
&0.20-0.35 & 0.23 & 0.24 & 6.1 & $-0.013\pm0.052\pm0.055$
& $0.136\pm0.074\pm0.022$ & $-0.091\pm0.069\pm0.039$ & 
$0.000\pm0.069\pm0.008$\\
\noalign{\smallskip}
\hline\noalign{\smallskip}
\multirow{6}{*}{\rotatebox{90}{\mbox{$Q^{2}$[GeV$^2$]}}} \hspace{0.01cm}
&1.0-1.4 & 0.09 & 0.05 & 1.2 & $-0.032\pm0.022\pm0.016$ 
& $0.077\pm0.033\pm0.037$ & $0.000\pm0.029\pm0.007$ & 
$-0.030\pm0.030\pm0.004$\\
&1.4-1.8 & 0.10 & 0.07 & 1.6 & $-0.050\pm0.026\pm0.017$ 
& $0.100\pm0.037\pm0.016$ & $0.000\pm0.034\pm0.018$ & 
$0.021\pm0.034\pm0.006$\\
&1.8-2.4 & 0.12 & 0.09 & 2.1 & $-0.031\pm0.023\pm0.017$ 
& $0.025\pm0.032\pm0.035$ & $-0.035\pm0.033\pm0.015$ & 
$0.074\pm0.034\pm0.009$\\
&2.4-3.2 & 0.14 & 0.11 & 2.8 & $-0.021\pm0.024\pm0.026$ 
& $0.106\pm0.038\pm0.013$ & $0.045\pm0.036\pm0.021$ & 
$-0.042\pm0.035\pm0.003$\\
&3.2-4.5 & 0.16 & 0.14 & 3.8 & $-0.010\pm0.027\pm0.014$
& $0.026\pm0.037\pm0.029$ & $0.018\pm0.037\pm0.015$ & 
$-0.003\pm0.037\pm0.004$\\
&4.5-10.0 & 0.23 & 0.20 & 5.8 & $-0.010\pm0.032\pm0.035$
& $0.055\pm0.046\pm0.023$ & $-0.095\pm0.046\pm0.025$ & 
$0.013\pm0.045\pm0.003$\\
\noalign{\smallskip}
\hline\noalign{\smallskip}
\end{tabular}
\end{center}

\begin{center}
\begin{tabular}{r|ccccrrr}
\hline\noalign{\smallskip}
\multicolumn{2}{c}{kinematic bin} 
&$\langle -t \rangle$ &$\langle x_N \rangle$ &$\langle Q^2
\rangle$ &$\ALUDVCS^{\sin \, \phi}$\hspace{0.6cm}
&$\ALUI^{\sin \, \phi}$\hspace{1.1cm} &$\ALUI^{\sin \, 
(2\phi)}$\hspace{0.9cm} \\
\multicolumn{2}{c}{} &[GeV$^2$] & &[GeV$^2$] &$\pm \rm {\delta_{stat}} \pm 
\rm {\delta_{syst}}$ &$\pm \rm {\delta_{stat}} \pm \rm {\delta_{syst}}$ 
&$\pm \rm {\delta_{stat}} \pm \rm {\delta_{syst}}$ \\ 
\noalign{\smallskip}
\hline\noalign{\smallskip}
\multicolumn{2}{c}{overall} & 0.13 & 0.10 & 2.5 & $-0.007\pm0.033 
\pm0.007$ & $-0.192\pm0.035\pm0.031$ & $0.073\pm0.031\pm0.012$\\
\noalign{\smallskip}
\hline\noalign{\smallskip}
\multirow{6}{*}{\rotatebox{90}{\mbox{$-t$[GeV$^2$]}}} \hspace{0.01cm}
&0.00-0.03 & 0.02 & 0.07 & 1.7 & $-0.042\pm0.074\pm0.011$ 
& $-0.296\pm0.104\pm0.006$ & $0.056\pm0.071\pm0.011$\\
&0.03-0.06 & 0.04 & 0.09 & 2.2 & $-0.101\pm0.077\pm0.013$ 
& $-0.167\pm0.084\pm0.008$ & $0.034\pm0.072\pm0.009$\\
&0.06-0.10 & 0.08 & 0.10 & 2.4 & $0.032\pm0.080\pm0.032$ 
& $-0.064\pm0.081\pm0.010$ & $0.114\pm0.076\pm0.032$\\
&0.10-0.20 & 0.14 & 0.11 & 2.7 & $0.018\pm0.068\pm0.009$ 
& $-0.215\pm0.071\pm0.016$ & $-0.022\pm0.065\pm0.013$\\
&0.20-0.35 & 0.26 & 0.12 & 3.1 & $0.095\pm0.087\pm0.009$
& $-0.286\pm0.095\pm0.008$ & $0.206\pm0.085\pm0.024$\\
&0.35-0.70 & 0.46 & 0.11 & 3.5 & $-0.029\pm0.118\pm0.035$
& $0.003\pm0.122\pm0.005$ & $0.133\pm0.124\pm0.030$\\
\noalign{\smallskip}
\hline\noalign{\smallskip}
\multirow{6}{*}{\rotatebox{90}{\mbox{$x_N$}}} \hspace{0.01cm}
&0.03-0.06 & 0.12 & 0.05 & 1.3 & $-0.007\pm0.064\pm0.021$ 
& $-0.197\pm0.083\pm0.061$ & $0.080\pm0.066\pm0.015$\\
&0.06-0.08 & 0.10 & 0.07 & 1.8 & $0.012\pm0.069\pm0.018$ 
& $-0.286\pm0.096\pm0.032$ & $0.084\pm0.067\pm0.014$\\
&0.08-0.10 & 0.11 & 0.09 & 2.3 & $0.041\pm0.080\pm0.025$  
& $-0.017\pm0.080\pm0.031$ & $-0.018\pm0.075\pm0.010$\\
&0.10-0.13 & 0.13 & 0.11 & 2.9 & $-0.056\pm0.084\pm0.033$ 
& $-0.212\pm0.090\pm0.023$ & $0.060\pm0.080\pm0.013$\\
&0.13-0.20 & 0.17 & 0.16 & 4.0 & $-0.109\pm0.090\pm0.037$
& $-0.189\pm0.093\pm0.020$ & $0.029\pm0.083\pm0.002$\\
&0.20-0.35 & 0.23 & 0.24 & 6.1 & $0.222\pm0.160\pm0.053$
& $-0.313\pm0.161\pm0.032$ & $0.444\pm0.163\pm0.032$\\
\noalign{\smallskip}
\hline\noalign{\smallskip}
\multirow{6}{*}{\rotatebox{90}{\mbox{$Q^{2}$[GeV$^2$]}}} \hspace{0.01cm}
&1.0-1.4 & 0.09 & 0.05 & 1.2 & $-0.028\pm0.068\pm0.035$ 
& $-0.208\pm0.082\pm0.060$ & $0.052\pm0.065\pm0.011$\\
&1.4-1.8 & 0.10 & 0.07 & 1.6 & $0.175\pm0.079\pm0.030$ 
& $-0.222\pm0.087\pm0.049$ & $0.127\pm0.077\pm0.019$\\
&1.8-2.4 & 0.12 & 0.09 & 2.1 & $-0.108\pm0.076\pm0.020$ 
& $-0.124\pm0.077\pm0.029$ & $-0.011\pm0.071\pm0.007$\\
&2.4-3.2 & 0.14 & 0.11 & 2.8 & $0.005\pm0.083\pm0.023$ 
& $-0.244\pm0.091\pm0.037$ & $0.054\pm0.077\pm0.010$\\
&3.2-4.5 & 0.16 & 0.14 & 3.8 & $-0.045\pm0.086\pm0.037$
& $-0.169\pm0.088\pm0.022$ & $0.119\pm0.082\pm0.005$\\
&4.5-10.0 & 0.23 & 0.20 & 5.8 & $-0.038\pm0.104\pm0.010$
& $-0.233\pm0.105\pm0.006$ & $0.166\pm0.100\pm0.006$\\
\noalign{\smallskip}
\hline\noalign{\smallskip}
\end{tabular}
\label{tb:table3}
\end{center}
\end{table}

The two hatched bands in Figs.~\ref{fig:bca_uu} and \ref{fig:bsad_uu} are 
theoretical calculations for the incoherent process, based on two 
different ans\"atze for modeling GPDs~\cite{Goeke:2001tz} in the VGG 
model~\cite{Vdhcode} (the coherent process will be considered in Section 
6.3.). In this model, a GPD is written as a double 
distribution~\cite{Mul94,Rad97} complemented by a 
D-term~\cite{Polyakov:1999gs,theor_bsa1}:
\begin{itemize}
\item In the `factorized ansatz' (VGG Fact.), the dependences on $t$ and 
$(x,\xi)$ are uncorrelated. The $t$ dependence is written in accordance 
with proton elastic form factors. The $(x,\xi)$ dependence  is based on 
double distributions~\cite{Mul94} constructed from ordinary PDFs 
complemented with a profile function that characterizes the strength of 
the $\xi$ dependence; in the limit $b \rightarrow \infty$ of the profile 
parameter $b$, the GPD is independent of $\xi$~\cite{Radyushkin:1998es}.  
Note that $b$ is a free parameter to be experimentally determined 
independently for valence and sea quarks.
\item The `Regge ansatz' (VGG Regge) implements entanglement of the $t$ 
de\-pendence of the GPD with its dependence on $x$ and $\xi$. This feature 
is inspired by the traditional interpretation of measurements of elastic 
diffractive processes in terms of Regge phenomenology~\cite{Goeke:2001tz}, 
and finds further support in more recent phenomenological 
considerations~\cite{Diehl:2004cx,Guidal:2004nd}. This ansatz for GPDs 
hence uses for the $t$ dependence of the double distributions a soft 
Regge-type parameterization $\propto |\xi|^{-\alpha(0) + \alpha ' \, |t|}$ 
with $\alpha ' =  0.8$\,GeV$^{-2}\dots 0.9$\,GeV$^{-2}$ for quarks.
\end{itemize}

Both theoretical calculations are averaged at the cross section level over 
incoherent processes on the proton and neutron in each kinematic bin. In 
both calculations the D-term is assigned the value zero. Earlier, it was found 
that inclusion of a D-term with any significant magnitude in the 
double-distribution model of Ref.~\cite{Vdhcode} employing several 
variants of Regge or factorized ans\"atze with any choice of profile 
parameters fails to describe the BCA amplitudes measured at HERMES on a 
hydrogen target~\cite{hermes_ttsa,hermes_bca_2006}. The theoretical bands 
in Figs.~\ref{fig:bca_uu} and \ref{fig:bsad_uu} correspond to the range of 
values of the asymmetry amplitudes obtained by varying the profile 
parameters $b_{\rm val}$ and $b_{\rm sea}$ between unity and infinity. The 
theoretical calculations based on the factorized ansatz fail to describe 
the $t$ dependence of $\ACU^{\cos (0\phi)}$ and $\ACU^{\cos \phi}$ as seen 
in Fig.~\ref{fig:bca_uu}. The calculations based on the Regge ansatz for 
GPDs are in good agreement with the $t$ dependence of the measured 
asymmetry amplitudes with respect to the beam charge $\ACU^{\cos(n\phi)}$. 
Both ans\"atze predict that $\ACU^{\cos \phi}$ decreases with increasing 
$x_N$, which is not seen in the data. Both ans\"atze undershoot the 
asymmetry amplitudes with  respect to the beam helicity 
$\ALUI^{\sin(n\phi)}$.
\begin{table}
\caption{Simulated fractional contributions for coherent and resonant 
processes in each kinematic bin.}
\small
\begin{center}
\begin{tabular}{c|crr}
\noalign{\smallskip}
\hline
\multicolumn{2}{c}{\hspace{0.2cm}kinematic bin} & \hspace{1.0cm}coherent & 
\hspace{1.0cm}resonant\\
\hline
\multicolumn{2}{c}{\hspace{0.2cm}overall} & 0.176 & 0.174 \\
\hline\noalign{\smallskip}
\multirow{6}{*}{\rotatebox{90}{\mbox{$-t$[GeV$^2$]}}} \hspace{0.01cm}
& 0.00 - 0.03 & 0.481 & 0.064 \\
& 0.03 - 0.06 & 0.256 & 0.110 \\
& 0.06 - 0.10 & 0.130 & 0.150 \\
& 0.10 - 0.20 & 0.053 & 0.206 \\
& 0.20 - 0.35 & 0.017 & 0.289 \\
& 0.35 - 0.70 & 0.005 & 0.387 \\
\noalign{\smallskip}
\hline\noalign{\smallskip}
\multirow{6}{*}{\rotatebox{90}{\mbox{$x_N$}}}
& 0.03 - 0.06 & 0.258 & 0.161 \\
& 0.06 - 0.08 & 0.214 & 0.160 \\
& 0.08 - 0.10 & 0.176 & 0.173 \\
& 0.10 - 0.13 & 0.127 & 0.184 \\
& 0.13 - 0.20 & 0.078 & 0.203 \\
& 0.20 - 0.35 & 0.032 & 0.198 \\
\noalign{\smallskip}
\hline\noalign{\smallskip}
\multirow{6}{*}{\rotatebox{90}{\mbox{$Q^{2}$[GeV$^2$]}}}
& 1.0 - 1.4 & 0.253 & 0.133 \\
& 1.4 - 1.8 & 0.209 & 0.154 \\
& 1.8 - 2.4 & 0.172 & 0.172 \\
& 2.4 - 3.2 & 0.150 & 0.193 \\
& 3.2 - 4.5 & 0.109 & 0.219 \\
& 4.5 - 10.0 & 0.055 & 0.237 \\
\noalign{\smallskip}
\hline\noalign{\smallskip}
\end{tabular}
\label{tb:table4}
\end{center}
\end{table}

\subsection{Comparison of the deuteron results with the HERMES 
results on beam-charge and beam-helicity asymmetries on the proton}
\label{subsec:comparison_p_d}
\begin{figure}[!tb]
\includegraphics[width=1\columnwidth]{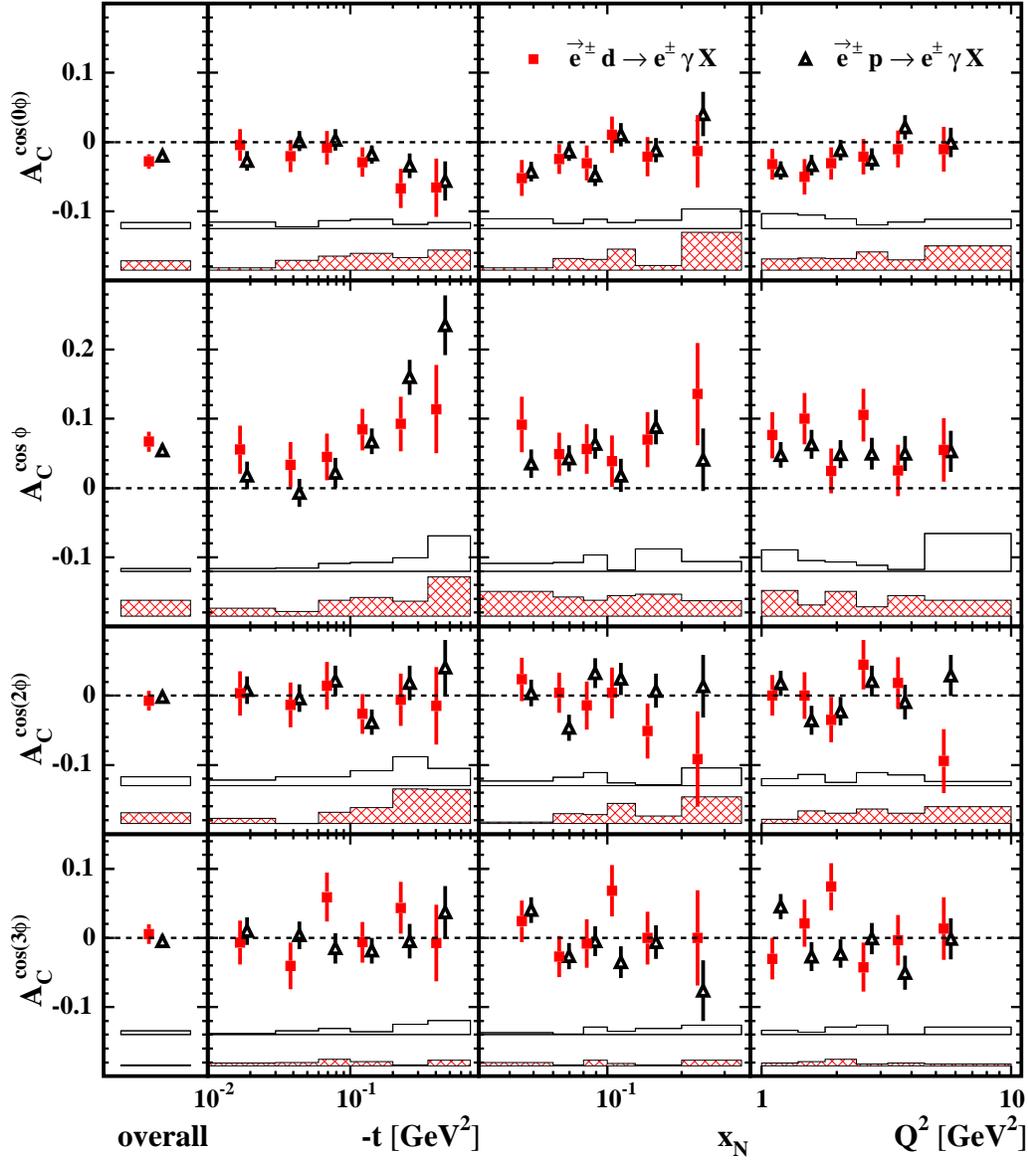}
\caption{Fourier amplitudes of the beam-charge asymmetry, which are 
sensitive to the interference term, in bins of $-t$, $x_N$, or $Q^2$, 
extracted from deuteron data (squares) and from proton data (triangles). 
The points for deuterium are slightly shifted along the x-axis for 
visibility. The error bars (bands) represent the statistical (systematic) 
uncertainties. The hatched band is for the deuterium target.}
\label{fig:bca_uu_pd}
\end{figure}

\begin{figure}[!tb]
\includegraphics[width=1\columnwidth]{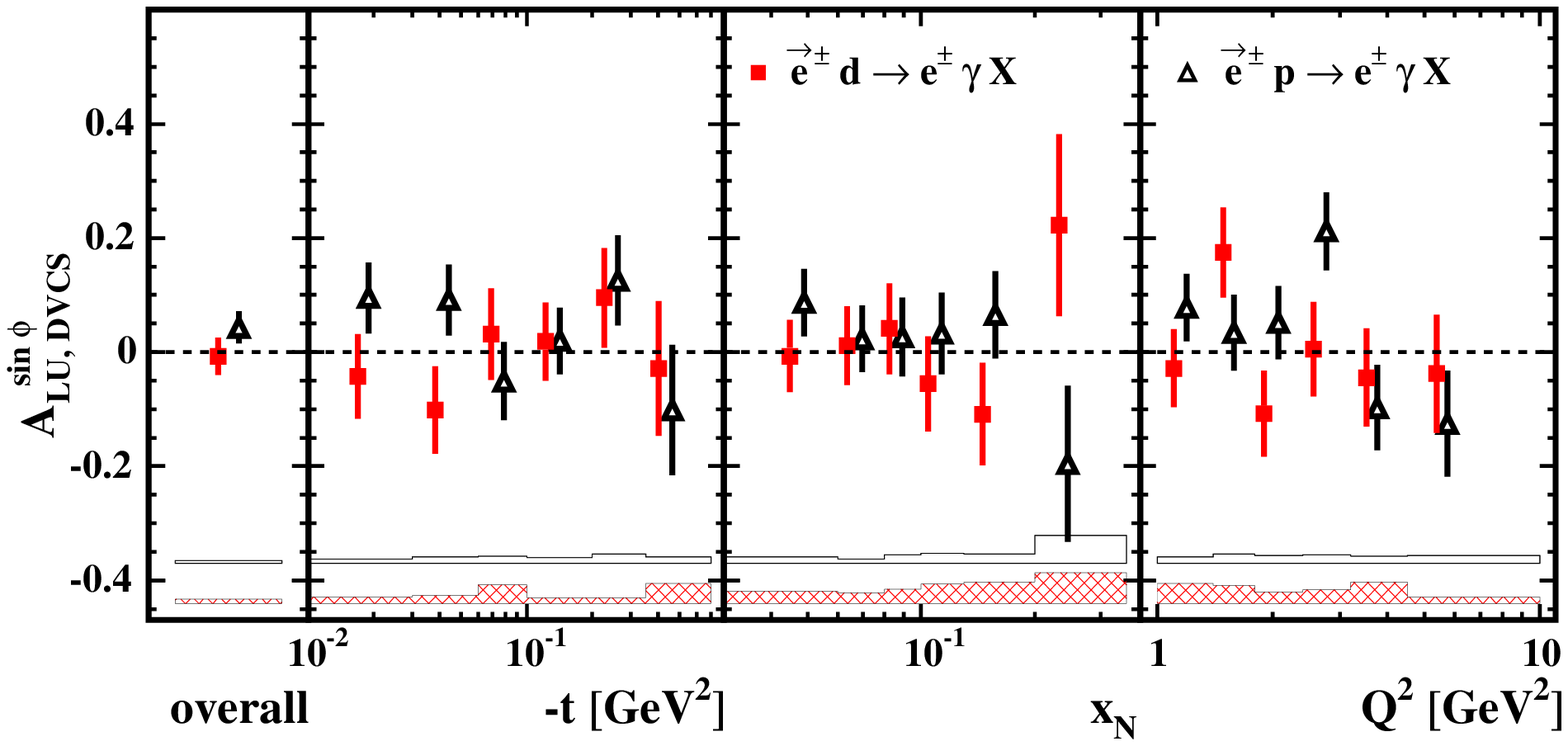}
\caption{Fourier amplitudes of the beam-helicity asymmetry that are 
sensitive to the squared DVCS term, in bins of $-t$, $x_N$, or $Q^2$, 
extracted from deuteron data (squares) and from proton data (triangles).
The error bars (bands) represent the statistical (systematic) 
uncertainties, which include all sources apart from the 2.4$\%$ 
(2.8$\%$) scale uncertainty for the deuteron (proton) data due to 
the beam polarization. The hatched band is for the deuterium target.}
\label{fig:bsad_uu_pd}
\end{figure}

\begin{figure}[!tb]
\includegraphics[width=1\columnwidth]{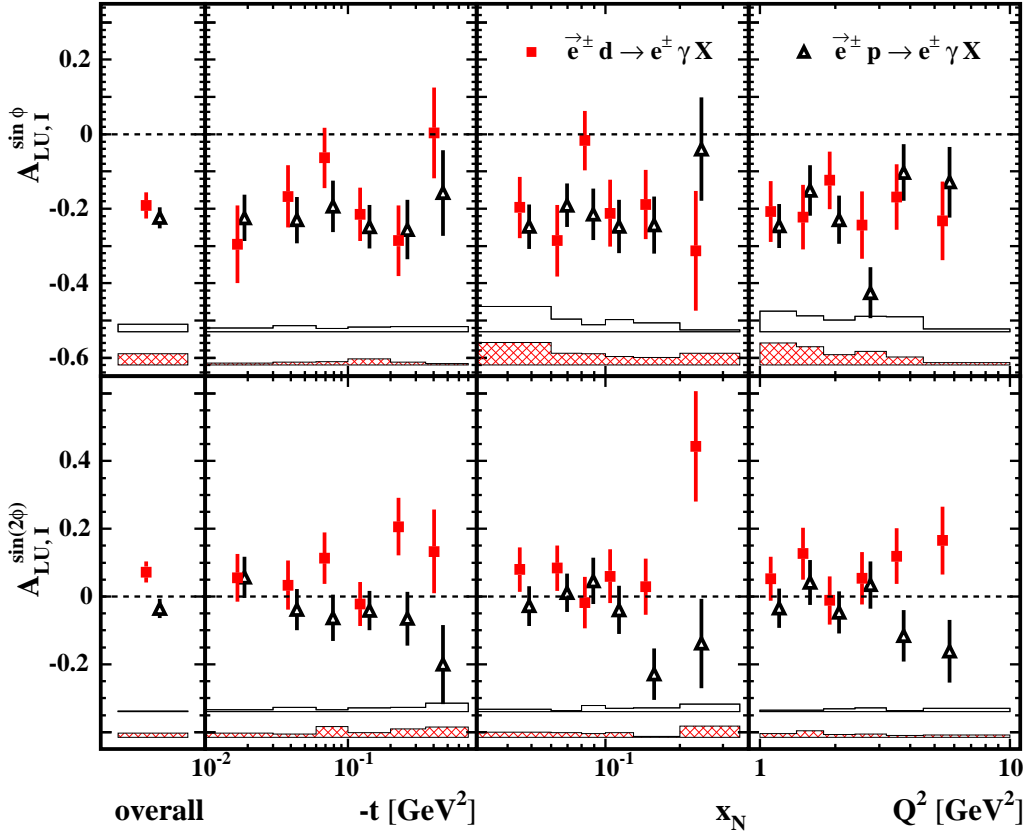}
\caption{Fourier amplitudes of the beam-helicity asymmetry that are 
sensitive to the interference term, in bins of $-t$, $x_N$, or $Q^2$, 
extracted from deuteron data (squares) and from proton data (triangles). 
The error bars (bands) represent the statistical (systematic) 
uncertainties, which include all sources apart from the 2.4$\%$ 
(2.8$\%$) scale uncertainty for the deuteron (proton) data due to 
the beam polarization. The hatched band is for the deuterium target.}
\label{fig:bsai_uu_pd}
\end{figure}

In Figs.~\ref{fig:bca_uu_pd}--\ref{fig:bsai_uu_pd} the overall 
asymmetry amplitudes as well as their $-t$, $x_N$, and $Q^2$ dependences, 
measured for the unpolarized deuterium target, are compared with the 
analogous results obtained from HERMES data on the 
proton~\cite{Proton_draft}.

The deuteron data include the coherent process $\vec e^{\, \pm}\,d \to 
e^{\pm}\,d\,\gamma$, and the incoherent process $\vec e^{\, \pm}\,d \to 
e^{\pm}\,p\,n\,\gamma$, where a nucleon may be excited to a resonance. The 
proton data include only $\vec e^{\, \pm}\,p \to e^{\pm}\,p\,\gamma$ and 
the case with resonance excitation. Any difference that appears at small 
values of $-t$ may be due to the coherent process. Monte Carlo simulations 
indicate that the incoherent process dominates for $0.06\rm\,GeV^2 <$ $-t$ 
$<\rm 0.7\rm\,GeV^2$ (see Fig.~\ref{figure1.2}). As shown in 
Figs.~\ref{fig:bca_uu_pd}-\ref{fig:bsai_uu_pd}, the deuteron and proton 
results are found to be consistent in most kinematic regions. A possible 
difference in the last two $-t$ bins of the amplitude $\ACU^{\cos \phi}$ 
(see Fig.~\ref{fig:bca_uu_pd}) may be due to the contributions of the 
neutron and its resonances. The proton and deuteron results for the 
amplitude $\ALUI^{\sin(2\phi)}$ integrated over the acceptance differ by 
2.5 times the total experimental uncertainties. This possible discrepancy 
is most evident at large $-t$ and large $x_N$ (or $Q^2$). Such a 
discrepancy would have no obvious explanation.

\subsection{Estimates of the asymmetries from coherent scattering}
\label{subsec:estimates_coh}
Estimates of the asymmetries for coherent scattering in the range $-t 
<\rm 0.06\rm\,GeV^2$, corresponding to the first two bins, were derived by 
correcting for the incoherent contributions of the proton and its 
resonances using the simulated fractional coherent contributions from 
Table~\ref{tb:table4}, under the assumption that the asymmetries for these 
contributions are the same as those on the free proton. The 
simulated contribution of approximately 7$\%$ from the process $\vec e^{\, 
\pm}\,n \to e^{\pm}\,n\,\gamma$ is estimated to have an effect on the 
asymmetries of less than 0.01. The extracted coherent asymmetries 
$\ACUcoh^{\cos \phi}$ and $\ALUcoh^{\sin \phi}$ are found to be 
$0.11 \pm 0.07\;(\mathrm{stat.}) \pm0.03\;(\mathrm{syst.})$ and $-0.29 \pm 
0.18\;(\mathrm{stat.}) \pm 0.03\;(\mathrm{syst.})$, respectively, at the 
average kinematic values\footnote{Nucleonic Bjorken $x_N$ is 
experimentally irrelevant for coherent scattering.} $\langle -t \rangle = 
0.03$ GeV$^2$, $\langle x_D \rangle = 0.04$, and $\langle Q^2 \rangle = 
1.9$ GeV$^2$. 

These results for the coherent asymmetries are compared in 
Table~\ref{tb:table5} with model estimates using the models $A$,  $B$,  
$B_0$, $\widehat B$, $B^\prime$, and $C$ of Refs.~\cite{DVCS2,theor_deu}, 
the main parameters of which are listed in Table~\ref{tb:table6}. The model 
estimates are based on the double distribution 
ansatz~\cite{Radyushkin:1998es} for nucleonic GPDs, combined with a 
factorized $t$ dependence, and with the D-term set to zero. The nucleonic 
GPDs are combined using the impulse approximation. The contribution of sea 
quarks is neglected in model $B_0$, while it is enhanced in model $C$ by a 
choice of a smaller value of the parameter $b_{\rm sea}$, which increases 
the absolute value of the beam-helicity asymmetry amplitude $\ALUcoh^{\sin 
\phi}$ compared to model $A$. In model $B^\prime$ ($\widehat B$), the GPD 
$H_3$ ($H_5$) is taken into account by arbitrarily equating it with $H_1\ 
(H_1(x)-H_1(-x))$. All other GPDs are kinematically suppressed and are set 
to zero. The models $B_0$ and $C$ were previously ruled out by the 
beam-helicity and beam-charge asymmetry measurements on the hydrogen 
target 
\cite{hermes_bsa:2001,CLAS_bsa:2001,hermes_ttsa,hermes_bca_2006,Proton_draft}.

Table~\ref{tb:table5} also includes model predictions from 
Ref.~\cite{theor_bsa1}. This model is based on double distributions, 
where only the polarizations of the valence quarks are considered 
for the nucleonic GPDs. A factorized ansatz for the $t$ dependence of the 
nucleonic GPDs is employed and the strange quark contribution is 
neglected. Again the impulse approximation is used to combine the 
nucleonic GPDs, without including the particular contribution from the 
D-term.

All models are consistent within two standard deviations in the 
total experimental uncertainty with the extracted results for 
$\ALUcoh^{\sin \phi}$ and $\ACUcoh^{\cos \phi}$, except for models $B_0$ 
and that of Ref.~\cite{theor_bsa1}, which disagree with the results of 
$\ACUcoh^{\cos \phi}$ by about 3.5 standard deviations. Here, it should be 
noted that predictions for the real part of the CFFs are subject to 
delicate cancellations~\cite{DVCS2} and hence are extremely sensitive to 
assumptions.

\begin{table}[!tb]
\begin{center}
\caption{Experimental and theoretical values of the beam-helicity and 
beam-charge asymmetries for the coherent process on the deuteron.
The theoretical predictions are for variants of the models of 
Ref.~\cite{DVCS2,theor_deu} and a model from Ref.~\cite{theor_bsa1}. The 
experimental uncertainties do not account for the model dependence of the 
simulated fractional contributions of coherent and incoherent processes.}
\begin{tabular}{crrrrrrrr}
\noalign{\smallskip}
\hline
& \multicolumn{1}{c}{Exp. value} & 
\multicolumn{7}{c}{Model} \\
\hline
& \hspace{0.2cm}value $\pm \, \rm {\delta_{stat}} \pm \rm {\delta_{syst}}$ 
& \hspace{0.8cm}A & \hspace{0.8cm}B & \hspace{0.8cm}${\rm B}_0$ & 
$\hspace{0.8cm}\widehat B$ & \hspace{0.8cm}$B^\prime$ & \hspace{0.8cm}C 
&\hspace{0.8cm}~\cite{theor_bsa1}\\
\hline
$\ALUcoh^{\sin \phi}$ & $-0.29\pm0.18\pms0.03$ &
-0.44 & -0.38  & -0.16 & -0.37 & -0.39 & -0.58 & -0.36\\
$\ACUcoh^{\cos \phi}$ & $\, 0.11\pm0.07\pm0.03$ &
0.10 & 0.09  & -0.17 & 0.09 & 0.09 & 0.22 & -0.15\\
\hline
\end{tabular}
\label{tb:table5}
\end{center}
\end{table}

\begin{table}
\caption{Model parameter sets for the GPD $H_1$ of the 
deuteron~\cite{DVCS2,theor_deu}. The $t$ slope parameter $B_{\rm 
sea}$ is used mainly to change the normalization of the sea quark GPD 
$H_1$.}
\begin{center}
\begin{tabular}{ccccc}
\noalign{\smallskip}
\hline
deuteron $H_1$ GPD & \multicolumn{4}{c}{Model} \\
\hline
Model parameters & \hspace{1.0cm}A & \hspace{1.0cm}B ($B^\prime$, 
$\widehat B$) & \hspace{1.0cm}${\rm B}_0$ & \hspace{1.0cm}C \\
\hline
$b_{\rm val}$ & \hspace{1.0cm}1 & \hspace{1.0cm}$\infty$ & 
\hspace{1.0cm}$\infty$ & \hspace{1.0cm}1 \\
$b_{\rm sea}$ & \hspace{1.0cm}$\infty$ & \hspace{1.0cm}$\infty$ & 
\hspace{1.0cm}$-$ & \hspace{1.0cm}1 \\
$B_{\rm sea}$ $[{\rm GeV}^{-2}]$ & \hspace{1.0cm}20 & \hspace{1.0cm}20 & 
\hspace{1.0cm}$-$ & \hspace{1.0cm}15 \\
\hline
\end{tabular}
\end{center}
\label{tb:table6}
\end{table}

\section{Summary}
\label{sec:summary}
Azimuthal asymmetries with respect to beam-helicity and beam-charge are 
measured for hard exclusive electroproduction of photons in deeply inelastic 
scattering off an unpolarized deuterium target. The observed asymmetries 
are attributed to  either the interference between the DVCS and the 
Bethe-Heitler processes or the pure DVCS process. The asymmetries are 
observed in the exclusive missing-mass domain $-(1.5)^2{\rm\,GeV^2} 
<  M_X^2 < (1.7)^2{\rm\,GeV}^2$. The dependences of these asymmetries on 
$-t$, $x_N$, or $Q^2$ are investigated. The results from the deuterium 
target include the coherent process $\vec e^{\;\pm}\,d \to 
e^{\pm}\,d\,\gamma$ and the incoherent process $\vec e^{\;\pm}\,d \to 
e^{\pm}\,p\,n\,\gamma$, where a nucleon may be excited to a resonance. For 
an unpolarized deuterium target, the leading Fourier amplitude of the 
beam-helicity asymmetry that is sensitive to the interference term is 
found to be substantial, but no significant $t$ dependence is observed. 
The leading amplitude of the beam-charge asymmetry is substantial at large 
$-t$, but becomes small at small values of $-t$. The amplitudes of the 
beam-helicity asymmetry that are sensitive to the squared DVCS term are 
found to be consistent with zero. The data are able to discriminate 
among various GPD models.

The measured asymmetry amplitudes from unpolarized deuteron and 
proton~\cite{Proton_draft} targets are consistent in most 
kinematic regions, except possibly for the leading amplitude of the 
beam-charge asymmetry in the last two $-t$ bins, and the `overall' value 
of $\ALU^{sin(2\phi)}$.
 
The beam-charge and beam-helicity asymmetry amplitudes for coherent 
scattering from the deuteron are extracted from the asymmetry amplitudes 
measured on unpolarized deuteron and proton targets. When compared to the 
GPD models of Refs.~\cite{DVCS2,theor_deu}, the results disfavor a large 
sea quark contribution while favoring a non-zero contribution. The results 
disfavor the variants of the model of Refs.~\cite{DVCS2,theor_deu} that 
omit sea quark contributions, and also the model of 
Ref.~\cite{theor_bsa1}.

\section{Acknowledgments}
\label{sec:Acknowledgments}
We gratefully acknowledge the DESY management for its support and the 
staff at DESY and the collaborating institutions for their significant 
effort. This work was supported by the FWO-Flanders and IWT, Belgium; 
the Natural Sciences and Engineering Research Council of Canada; the 
National Natural Science Foundation of China; the Alexander von Humboldt 
Stiftung; the German Bundesministerium f\"ur Bildung und Forschung (BMBF); 
the Deutsche Forschungsgemeinschaft (DFG); the Italian Istituto Nazionale 
di Fisica Nucleare (INFN); the MEXT, JSPS, and G-COE of Japan; the Dutch 
Foundation for Fundamenteel Onderzoek der Materie (FOM); the 
U.K.~Engineering and Physical Sciences Research Council, the Science and 
Technology Facilities Council, and the Scottish Universities Physics 
Alliance; the U.S.~Department of Energy (DOE) and the National Science 
Foundation (NSF); the Russian Academy of Science and the Russian Federal 
Agency for Science and Innovations; the Ministry of Economy and the 
Ministry of Education and Science of Armenia; and the European 
Community-Research Infrastructure Activity under the FP6 ``Structuring the 
European Research Area" program (HadronPhysics, contract number 
RII3-CT-2004-506078).

\end{document}